\documentclass[journal=jacsat,manuscript=article]{achemso}

\SectionNumbersOn

\usepackage[version=3]{mhchem} 
\usepackage{siunitx}      
\usepackage{booktabs}
\usepackage{graphicx}
\usepackage[utf8]{inputenc}
\DeclareUnicodeCharacter{2009}{ } 
\usepackage[utf8]{inputenc}
\author{Mehedi Hasan}
\affiliation[]{Department of Industrial and Production Engineering, Bangladesh University of Engineering and Technology, Dhaka--1000, Bangladesh}
\email{mehedihasanbuet29@gmail.com}

\author{Ishtiaq Murshed}
\affiliation[]{Department of Chemical Engineering, Bangladesh University of Engineering and Technology, Dhaka--1000, Bangladesh}

\author{Khayrul Islam}
\affiliation[]{Department of Mechanical Engineering, Lehigh University, Bethlehem, PA 18015, USA}

\author{A.K.M. Masud}
\affiliation[Bangladesh University of Engineering and Technology]
{Department of Industrial and Production Engineering, Bangladesh University of Engineering and Technology, Dhaka, Bangladesh}

\title[An \textsf{achemso} demo]
{Toward Unified Interphase Engineering: The Solid–Electrolyte Interphase in Batteries and Supercapacitors}
\abbreviations{SEI,EDLC,XPS,FTIR,EQCM,EIS,DFT,MD,ML,ESR}
\keywords{solid electrolyte interphase, supercapacitors, batteries, electric double layer, interface engineering, multiscale modeling}

\begin{document}

\section*{Abstract}
The development of next-generation electrochemical energy storage requires devices that synergistically combine the high energy density of batteries with the exceptional power capability and cycle life of supercacitors, yet fundamental understanding of the interfacial phenomena governing performance across these platforms remains fragmented. While the solid–electrolyte interphase (SEI), a nanometer-scale passivation layer formed by electrolyte decomposition, has been extensively characterized in battery systems, analogous interfacial films in supercapacitors have received limited systematic investigation despite mounting experimental evidence for their existence and functional significance. This review advances the thesis that SEI formation constitutes a universal electrochemical phenomenon arising whenever applied potentials drive electrode Fermi levels into electrolyte molecular orbitals beyond thermodynamic stability limits, independent of whether charge storage proceeds via Faradaic redox or non-Faradaic electrostatic mechanisms. Distinctions between battery SEIs (10–100 nm thickness, quasi-static evolution) and supercapacitor interphases (1–10 nm, dynamic reconstruction) reflect operational boundary conditions (potential range, cycling frequency, and ion flux) rather than fundamental mechanistic divergence. Quantitative analysis reveals that ionic resistivity scales universally with inorganic-phase volume fraction across device architectures, while self-limiting growth follows identical electron-tunneling-mediated passivation regardless of platform. Critically, rationally engineered interphases achieved through electrolyte additive optimization, atomic-layer-deposited protective coatings, or electrode surface functionalization suppress parasitic leakage currents by 60–80\%, maintain capacitance retention above 95\% through 50,000 cycles, and enable stable operation exceeding 3.0 V in organic and ionic-liquid electrolytes. By establishing shared mechanistic principles and facilitating systematic knowledge transfer between battery and supercapacitor research communities, this unified framework enables predictive interphase engineering strategies that deliver battery-level energy density, capacitor-level power capability, and ultralong operational lifetimes for electrified transportation, renewable grid integration, and sustainable infrastructure.
\section{Introduction}

The accelerating global transition toward electrification and renewable energy integration has driven the relentless pursuit of advanced electrochemical energy-storage technologies~\cite{detka2023selected,navarro2021present}. 
From smartphones and laptops to electric vehicles and grid-scale buffering, the demand for reliable, efficient, and sustainable energy-storage devices continues to rise~\cite{whittingham2014ultimate,goodenough2013li}. 
At the core of this technological revolution lie two complementary families of systems: rechargeable batteries and supercapacitors, each occupying a distinct yet overlapping domain of the energy–power landscape~\cite{beyers2023ragone,burke2000ultracapacitors}. 
Batteries, exemplified by lithium-ion and sodium-ion chemistries, deliver high energy density (100–300~Wh~kg$^{-1}$) through bulk Faradaic redox reactions, while supercapacitors, or electrochemical capacitors, provide rapid charge–discharge capability (up to 10~kW~kg$^{-1}$) and ultralong cycle life (exceeding $10^6$ cycles) through surface-controlled charge accumulation~\cite{ahn2024strategies,lin2021prospect,choi2020achieving}. 
The synergistic coexistence of these devices forms the backbone of modern electrochemical energy storage, and the quest to merge their advantages has sparked immense scientific and industrial interest~\cite{li2020recent,wang2016electrochemical}.

\begin{figure}[htbp]
  \centering
\includegraphics[width=0.9\textwidth,page=1]{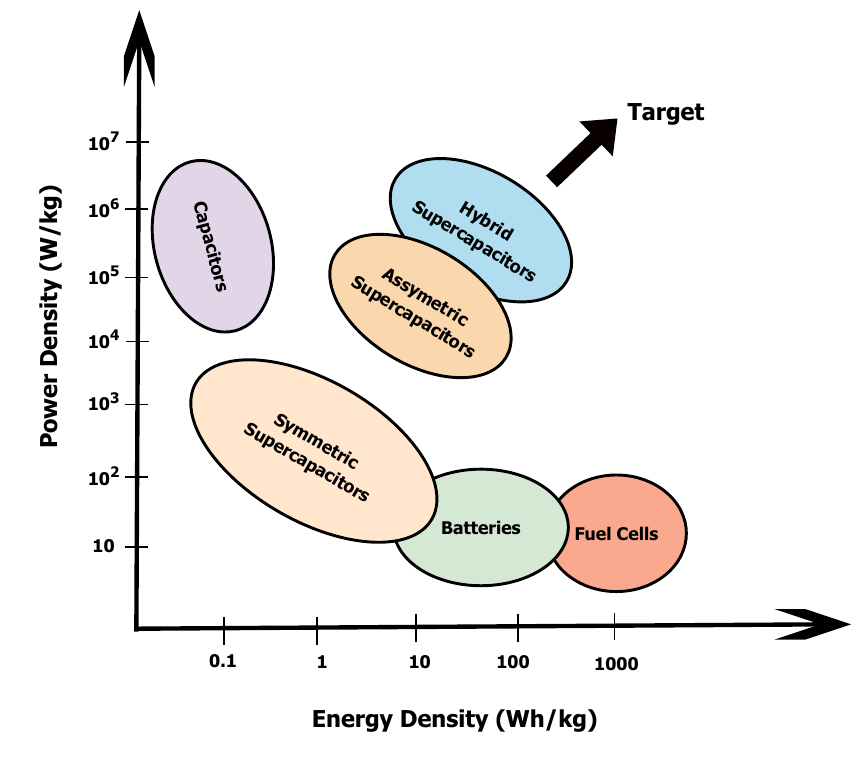}
\caption{The Electrochemical Energy–Power Landscape.}
  \label{fig:Intro_1}
\end{figure}

The concept of an interfacial passivation layer in electrochemical systems dates back to the pioneering work of Emanuel Peled in 1979, who first described the solid–electrolyte interphase (SEI) on alkali-metal electrodes~\cite{peled1979electrochemical}. 
Peled observed that a spontaneously formed, nanometer-scale film (composed of electrolyte decomposition products) electronically insulates the electrode while permitting selective ion transport, thereby preventing continuous electrolyte breakdown~\cite{peled2017sei}. 
This discovery revolutionized battery science, establishing the SEI as an indispensable component governing cycle life, Coulombic efficiency, and safety in lithium-ion, sodium-ion, and lithium-metal batteries~\cite{adenusi2023lithium,an2016state,wu2021recent}.

Over the ensuing decades, intensive research effort has been directed toward elucidating SEI formation mechanisms, chemical composition, morphology, and dynamic evolution in rechargeable batteries~\cite{cheng2016review,tan2023structural,wang2018review}. 
Techniques such as X-ray photoelectron spectroscopy (XPS), cryogenic transmission electron microscopy (cryo-TEM), and operando electrochemical quartz-crystal microbalance (EQCM) have revealed that battery SEIs are typically 10–100~nm thick, composed of inorganic salts (e.g., LiF, Li$_2$CO$_3$, Li$_2$O) in the inner region and polymeric organic species (e.g., polycarbonates, alkoxides) toward the outer surface~\cite{owejan2012solid,huang2019evolution,kitz2018operando}. 
These layers evolve slowly over thousands of cycles and must endure significant mechanical stress due to volumetric changes in active materials during lithiation–delithiation~\cite{d_2021}.

In contrast, the interfacial chemistry of supercapacitors has long been assumed to be purely non-Faradaic, dominated by electrostatic ion adsorption at the electric double layer (EDL) without chemical bond formation or electrolyte decomposition~\cite{simon2008materials,salanne2016efficient}. 
However, recent high-resolution spectroscopic and microscopic investigations have challenged this assumption, revealing the presence of ultrathin (1–10~nm), dynamic SEI-like films on activated-carbon, graphene, and MXene electrodes cycled at potentials above 2.5–3.0~V in organic, ionic-liquid, and even aqueous "water-in-salt" electrolytes~\cite{frackowiak2013electrode,quan2021unveiling,yvenat2023study}. 
These films share compositional similarities with battery SEIs (including inorganic fluorides, carbonates, and polymeric species) but exhibit greater structural reversibility, elasticity, and continuous reconstruction under rapid charge–discharge cycling~\cite{pamete2023many,hwang2019fundamental,ding2020ultra}.

Despite mounting experimental evidence for SEI formation in supercapacitors, the phenomenon remains poorly understood and lacks a unified theoretical framework. 
Several critical questions remain unanswered: 
(i)~What are the thermodynamic and kinetic conditions that trigger SEI nucleation in non-Faradaic systems? 
(ii)~How do the composition, thickness, and morphology of supercapacitor SEIs differ from those in batteries, and what governs these differences? 
(iii)~Can mechanistic insights and engineering strategies developed for battery SEIs (such as electrolyte additive design, artificial interphase coatings, and controlled formation protocols) be systematically translated to supercapacitors? 
(iv)~How does the dynamic, self-healing nature of supercapacitor SEIs influence long-term cycling stability, self-discharge, and power performance?

\begin{figure}[htbp]
  \centering
\includegraphics[width=0.9\textwidth,page=1]{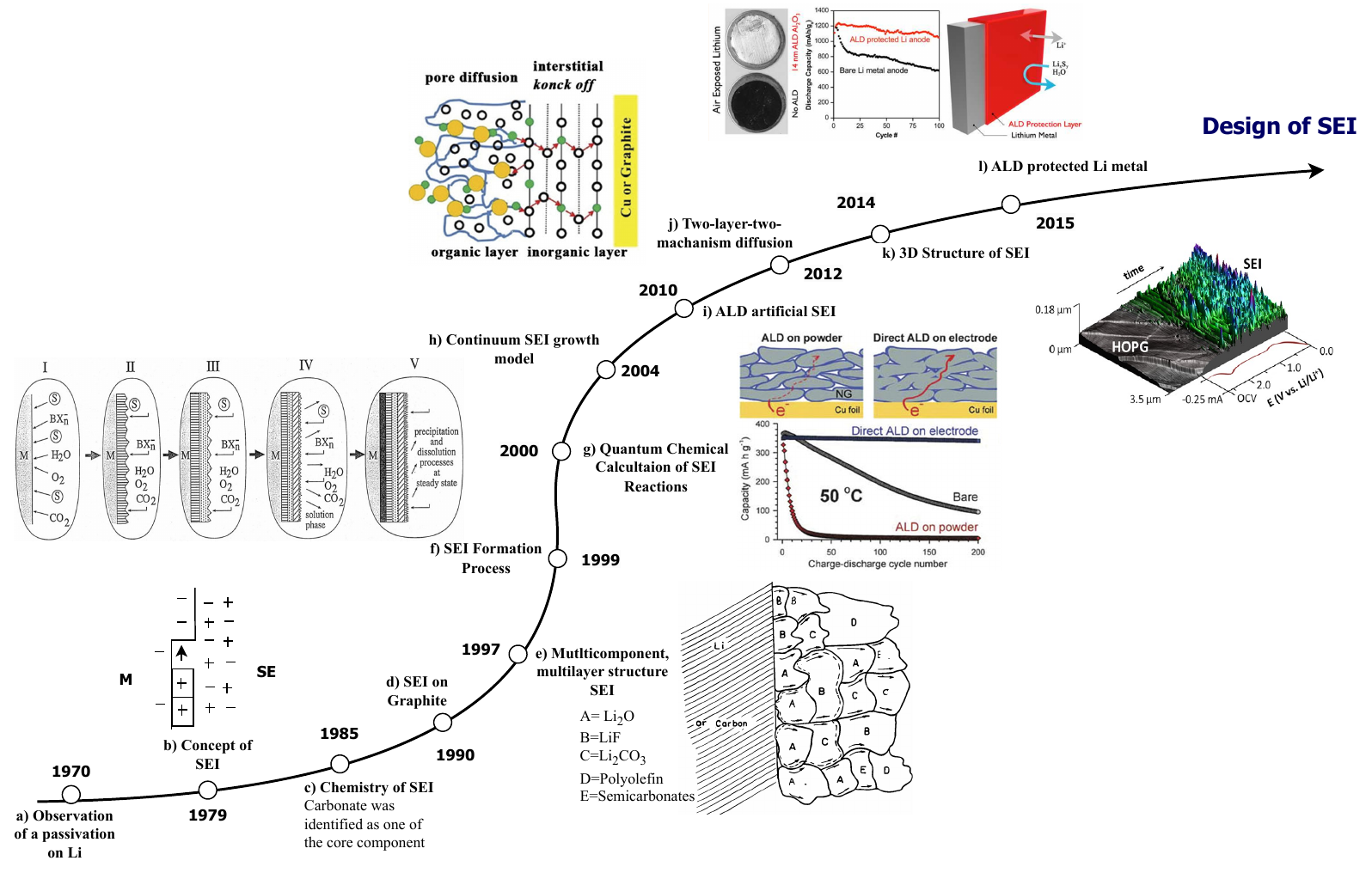}
\caption{The historical of solid electrolyte interphase.}
  \label{fig:Intro_history}
\end{figure}

The history of the Solid Electrolyte Interphase (SEI) on negative electrodes in lithium-ion batteries spans over four decades, evolving from initial observation to sophisticated design. The journey began in the early 1970s when Dey first observed a passivation layer on lithium metal in Figure-2(a)\cite{dey1970film}. In 1979, Peled\cite{peled1979electrochemical}  formally introduced the concept of the SEI  Figure-2(b). Subsequent experimental work focused on the SEI's chemistry, with Nazri and Muller(1985) and Aurbach et al.(1987)\cite{agubra2014formation} identifying $\text{Li}_2\text{CO}_3$ as a main component Figure-2(d,e). The effective SEI layer on graphite was confirmed in 1990 (b)\cite{fong1990studies}. This understanding led Peled\cite{peled1997advanced} to conceptualize the SEI as a mosaic structure and translate it into an equivalent circuit model in 1997 Figure-2(e), while Aurbach et al.\cite{aurbach1999new} illustrated the formation process starting from electrolyte reduction Figure-2(d). The early 2000s saw the use of quantum chemical calculations to simulate the reaction pathways contributing to SEI formation Figure-2(g)\cite{wang2001theoretical}, followed by physics-based continuum models developed in 2004 to simulate SEI growth by Christensen et al\cite{christensen2004mathematical}.  This deeper understanding culminated in the two-layer/two-mechanism model proposed by Shi et al.\cite{shi2012direct}, based on calculated Li-ion diffusion in the $\text{Li}_2\text{CO}_3$ inorganic layer in 2012 Figure-2(k). Direct observation of the multi-component and multi-layer SEI formation over time was achieved by Cresce et al.\cite{cresce2014situ} in 2014 using in situ electrochemical AFM. The fundamental understanding of SEI properties and formation has since led to the design of artificial SEIs using techniques like atomic layer deposition (ALD), with Jung et al.(2015)\cite{jung2010ultrathin} demonstrating improved durability on graphite and Kozen et al.(2015)\cite{kozen2015next} using ALD to protect lithium metal Figure-2(l).

Current literature on supercapacitor interphases is fragmented across materials science, electrochemistry, and device engineering, with limited cross-referencing to the extensive body of battery SEI research~\cite{lim2023advances,pathak2024high,patel2024review}. 
Moreover, the terminology is inconsistent: some authors refer to "interfacial films," "surface layers," or "electrolyte decomposition products" without explicitly invoking the SEI concept~\cite{yuan2022influencing,xu2025recent}. 
This conceptual disconnect has hindered the development of predictive models, unified characterization protocols, and transferable design principles that could accelerate supercapacitor performance optimization~\cite{yao2025roadmap}.

This review advances the thesis that SEI formation is a universal electrochemical phenomenon arising whenever the applied potential drives the electrode–electrolyte system beyond the thermodynamic stability window of the electrolyte, independent of whether the primary charge-storage mechanism is Faradaic (batteries) or non-Faradaic (supercapacitors)~\cite{suo2017solid,spotte2022towards,adenusi2023lithium}. 
The distinctions between battery and supercapacitor SEIs (thickness, composition, dynamics, and mechanical properties) reflect differences in operating conditions (potential range, cycling frequency, ion flux) rather than fundamental differences in formation mechanisms~\cite{wang2018review,an2016state}.

At the molecular level, SEI nucleation begins when the electrode Fermi level intersects the lowest unoccupied molecular orbital (LUMO) of the electrolyte, enabling electron tunneling into solvated molecules and initiating bond cleavage~\cite{li2023review,madhusudhanan2023revisiting}. 
This process triggers a cascade of reactions (solvent reduction, salt decomposition, and radical polymerization) that produce insoluble fragments nucleating on the electrode surface~\cite{wang2020reductive,banerjee2025formation}. 
The resulting interphase comprises a heterogeneous mixture of inorganic (e.g., LiF, Li$_2$CO$_3$, BF$_x$O$_y$) and organic (e.g., polycarbonates, nitrile oligomers) species arranged in a bilayer architecture~\cite{yan2021nucleation,wu2025revealing}.

Crucially, this self-limiting growth mechanism (wherein the interphase thickens until electron tunneling becomes negligible) operates identically in both batteries and supercapacitors, differing only in the final equilibrium thickness and the rate of dynamic reconstruction~\cite{peak1972diffusion,ramadass2002capacity}. 
By recognizing this mechanistic continuity, we can systematically transfer knowledge, experimental techniques, and engineering strategies between these traditionally separate research domains~\cite{warburton2021tailoring,liu2024exploring}.

This review synthesizes fundamental principles, experimental characterization, computational modeling, and engineering strategies for SEI control across batteries and supercapacitors, progressing systematically from atomic-scale mechanisms to device-level performance optimization.

This review synthesizes knowledge from multiple research communities (electrochemistry, materials science, battery engineering, and supercacitor technology) to construct a unified understanding of SEI phenomena. 
Our literature search encompassed peer-reviewed journal articles, conference proceedings, and authoritative reviews published between 1979 (Peled's seminal SEI discovery) and 2025. 
Primary databases included Web of Science, Scopus, PubMed, Google Scholar, and specialized electrochemical databases. 
Search terms combined "solid electrolyte interphase," "SEI," "supercapacitor," "electrochemical capacitor," "EDLC," "pseudocapacitor," "interfacial film," "passivation layer," and "electrolyte decomposition" with Boolean operators to capture both battery-focused and capacitor-focused literature.

Inclusion criteria prioritized: 
(i)~experimental studies employing advanced characterization techniques (XPS, cryo-TEM, operando spectroscopy) to probe interfacial chemistry; 
(ii)~computational investigations using ab-initio methods, molecular dynamics, or machine learning to elucidate SEI mechanisms; 
(iii)~engineering studies demonstrating electrolyte additive effects, surface modifications, or artificial interphase strategies; and 
(iv)~review articles synthesizing battery SEI knowledge with potential applicability to supercapacitors. 
We excluded purely device-performance studies lacking mechanistic insight into interfacial processes and preliminary conference abstracts without peer-reviewed follow-up.

Cross-referencing was performed to identify seminal works cited across both battery and supercapacitor communities, revealing key conceptual bridges (e.g., water-in-salt electrolytes, film-forming additives, operando EQCM) that enable knowledge transfer between domains. 
Over 400 references are integrated into this review, organized thematically to guide readers from fundamental principles through cutting-edge applications.

Establishing SEI formation as a universal phenomenon has profound implications for next-generation energy-storage device design. 
First, it enables predictive engineering: understanding how electrolyte composition, electrode surface chemistry, and operating conditions determine interphase properties allows rational design of SEIs with tailored thickness, ionic conductivity, and mechanical resilience~\cite{wan2024designing,jiang2024engineering}. 
Second, it facilitates cross-platform innovation: strategies proven successful in batteries (such as FEC or VC additives that form uniform fluorinated SEIs~\cite{yohannes2019sei,kitz2020operando}) can be systematically adapted to stabilize high-voltage supercapacitors. 
Conversely, the highly dynamic, self-healing SEIs observed in ionic-liquid capacitors offer insights for designing adaptive interfaces in next-generation battery chemistries~\cite{jankowski2019functional,yang2023characterization}.

Third, unified SEI frameworks accelerate multiscale modeling integration. 
Atomistic simulations (DFT, reactive MD) capture early-stage nucleation and chemical composition; continuum transport models describe long-term growth and impedance evolution; and machine-learning surrogates bridge these scales to predict device-level performance from molecular descriptors~\cite{du2025atomistic,kim2022multiscale,you2024principal}. 
Such hierarchical workflows, validated against operando experimental data, will enable in-silico optimization of electrolyte formulations and electrode architectures before costly laboratory synthesis~\cite{chen2020electrolyte,liu2024exploring}.

Fourth, recognizing the dynamic nature of supercapacitor SEIs (thinner, more elastic, continuously reconstructing) informs durability and reliability engineering. 
Unlike thick, quasi-static battery SEIs that crack under mechanical stress, supercapacitor interphases must tolerate millions of rapid charge–discharge cycles without catastrophic failure~\cite{pamete2023many,d_2021}. 
Designing cross-linked binder networks, self-healing polymer matrices, and resilient artificial coatings becomes paramount for achieving ultralong cycle life~\cite{ju2024self,liu2016artificial}.

Finally, this unified perspective supports sustainability and circular-economy goals. 
By understanding which SEI components are environmentally benign, biodegradable, or recyclable, researchers can design electrolytes and electrode surface treatments that minimize environmental footprint while maintaining performance~\cite{ng2020non,huang2023salt}. 
Moreover, predictive SEI models reduce trial-and-error experimentation, lowering resource consumption and accelerating time-to-market for next-generation supercapacitors powering electric vehicles, grid storage, and portable electronics~\cite{yao2025roadmap}.

\section{Fundamentals of the Solid–Electrolyte Interphase (SEI) in Batteries and Supercapacitors}

The concept of the solid–electrolyte interphase (SEI) originated in battery science, where electrolyte reduction produces a self-passivating layer that stabilizes electrode–electrolyte contact during cycling. Recent evidence suggests that analogous, albeit thinner and more dynamic, SEI-like films can also arise in supercapacitors under extreme potentials, high surface reactivity, or defect-rich conditions. Understanding these parallels allows cross-fertilization of ideas: mechanistic principles from batteries clarify interphase chemistry in capacitors, while the reversible, elastic behavior of capacitor interfaces offers new insight into SEI adaptability in batteries.

\subsection{Capacitive Charge Storage Fundamentals}

Supercapacitors (also known as electrochemical capacitors) store energy through fast and reversible interfacial processes that bridge the gap between conventional capacitors and batteries. Depending on the dominant charge-storage mechanism, they are broadly classified as electric double-layer capacitors (EDLCs) and pseudocapacitors. In EDLCs, charge storage is non-Faradaic and arises from electrostatic ion adsorption at the electrode–electrolyte interface. When an external potential is applied, oppositely charged ions accumulate near the electrode surface, forming an electric double layer (EDL) that acts as a nanoscale capacitor. The total stored charge depends on the accessible surface area of the electrode, the dielectric constant of the electrolyte, and the effective thickness of the EDL~\cite{sikiru2025projection, soshi_2019}. This mechanism provides excellent power density and long cycle life, as it does not involve structural changes or phase transformations in the electrode material.

Pseudocapacitors, by contrast, store charge through fast surface or near-surface Faradaic reactions that involve electron transfer between the electrode and electrolyte species. Transition-metal oxides (e.g., MnO$_2$, RuO$_2$) and conducting polymers (e.g., polyaniline, polypyrrole) are typical pseudocapacitive materials. The capacitance arises from reversible redox, intercalation, or adsorption processes that occur over a continuum of potentials rather than discrete battery-like plateaus. Because these reactions are surface-confined, pseudocapacitors combine the high energy densities of batteries with the power performance of EDLCs~\cite{lin2021prospect, tale2024novel, zhu2024preparation}. Hybrid or asymmetric supercapacitors integrate both mechanisms by pairing an EDLC-type electrode with a pseudocapacitive one, thereby expanding the voltage window and enhancing overall energy storage. The interfacial charge distribution that governs these processes also defines the potential gradients that initiate electrolyte decomposition and SEI nucleation in both batteries and capacitors.

\subsection{Interfacial Electric Field Distribution and Ion Dynamics}

The electric double layer governs the microscopic origin of capacitance in supercapacitors. At the electrode–electrolyte interface, ions rearrange in response to the surface potential, producing a space-charge region characterized by the Stern–Gouy–Chapman model~\cite{stern1924theorie, gouy1910constitution, chapman1913li}. The inner Helmholtz plane (IHP) consists of specifically adsorbed, partially desolvated ions in direct contact with the electrode surface, while the outer Helmholtz plane (OHP) contains solvated ions whose centers are separated by one solvent-molecule thickness~\cite{sikiru2025projection, dong2024nature}. Beyond these layers lies the diffuse region, where ion distribution follows a Boltzmann potential profile. The differential capacitance depends on the local dielectric constant, ion concentration, and potential-dependent surface charge density~\cite{sikiru2025projection}. In nanoporous electrodes, confinement effects alter ion packing and solvation structure, giving rise to non-classical phenomena such as ion desolvation, crowding, and overscreening~\cite{ge2025advanced}. These effects strongly influence voltage-dependent capacitance and energy density.

Spectroscopic, scattering, and molecular-dynamics studies reveal that the EDL is not a simple continuum but a dynamically fluctuating interfacial region with heterogeneous ion orientations and time-dependent correlations~\cite{tan2022nanoconfined}. The interplay of ion–ion and ion–solvent interactions determines the local permittivity and thus the capacitance. In pseudocapacitive materials, surface redox reactions further modify the interfacial electric field. These same electrostatic gradients set the stage for SEI formation once the local potential exceeds the electrolyte stability limit, linking double-layer physics with interphase chemistry across both energy-storage platforms.

\subsection{Origin and Function of the Solid–Electrolyte Interphase}

The electrode–electrolyte interface is a chemically complex region where ion adsorption, charge transfer, and molecular decomposition occur simultaneously. While EDL formation governs the fast and reversible electrostatic storage that defines supercapacitors, parasitic reactions at the same interface can lead to the emergence of a thin, passivating SEI. Much of the mechanistic framework for interpreting such interphases derives from battery research, where SEI formation is intrinsic and indispensable. In supercapacitors, similar reactions occur only under overstress or at defect-rich sites, producing a more transient and reversible film.

The SEI originates from partial electrochemical decomposition of electrolyte components—solvent molecules, salt anions, or additives—at electrode surfaces under high electric fields. When the applied potential exceeds the electrolyte’s stability window, redox-active species decompose to form insoluble fragments that nucleate on the surface~\cite{lim2023advances}. The resulting layer comprises a heterogeneous mixture of inorganic and organic compounds such as carbonates, fluorides, oxides, and polymerized species~\cite{yvenat2023study}. In batteries this process is deliberately exploited to stabilize electrodes; in capacitors it occurs incidentally yet can still enhance performance by suppressing leakage.

Unlike the purely capacitive EDL, which forms and dissipates reversibly, the SEI is a semi-permanent interphase that evolves during early cycles and stabilizes thereafter. Its dual function—electronically insulating while ionically conductive—renders it a self-passivating charge-transfer mediator. In batteries this duality enables stable Li$^+$ transport through tens-of-nanometer layers; in capacitors it operates over only a few nanometers, permitting rapid ionic exchange while limiting parasitic reduction. A well-formed SEI suppresses continuous decomposition, lowering leakage current and improving Coulombic efficiency~\cite{yao2025roadmap, patel2024review, hao2025long}. However, excessive growth can increase series resistance and slow kinetics, emphasizing the need for balanced composition and thickness control.

\subsection{Compositional and Structural Distinctions Across Energy-Storage Platforms}

Bridging SEI behavior across these two systems clarifies how shared electrochemical principles yield different interphase dynamics. In rechargeable batteries, the SEI is essential for long-term operation: it passivates the anode against continuous electrolyte decomposition while permitting Li$^+$ or Na$^+$ transport. Battery SEIs are typically tens of nanometers thick, composed of inorganic salts (LiF, Li$_2$O, Li$_2$CO$_3$) in the inner region and polymeric organic components (polycarbonates, alkoxides) toward the outer surface~\cite{adenusi2023lithium, grill2024long}. Once formed, these layers remain largely static and evolve slowly over many cycles, occasionally thickening due to secondary reactions.

In contrast, supercapacitor SEIs are thinner (often below 5–10~nm) and more dynamic because charge storage occurs primarily through non-Faradaic or surface-confined pseudocapacitive processes~\cite{akhter2023mxenes}. Interfacial potential fluctuations during rapid charging and discharging continually perturb the equilibrium structure, leading to reversible rearrangements or partial dissolution. Consequently, the SEI in supercapacitors behaves as a quasi-elastic interface that expands and contracts without catastrophic failure~\cite{zhao2024technological}. Furthermore, since the operating voltage of supercapacitors (1–3~V) is lower than typical battery cutoff voltages, electrolyte decomposition is limited, and SEI formation depends strongly on surface defects, residual moisture, and specific ion–surface interactions~\cite{adenusi2023lithium}.

From a compositional standpoint, organic-electrolyte supercapacitors often yield SEIs containing solvent fragments (acetonitrile or propylene carbonate) mixed with anion-derived fluorinated or borate species (BF$_4^-$, PF$_6^-$)~\cite{huang2023anion, tan2023structural}. In aqueous systems, decomposition of water forms hydroxide or oxide layers that are less stable and reconstruct during cycling, whereas ionic-liquid-based supercapacitors generate fluoride- or sulfur-enriched SEIs from anion cleavage~\cite{huang2023anion}. These dynamic interphases, although thinner, exhibit higher heterogeneity, influencing self-discharge and capacitance retention.

Mechanically, battery SEIs must accommodate large volumetric expansion (e.g., in Li metal or Si anodes), whereas supercapacitor electrodes—typically carbonaceous or oxide surfaces—undergo minimal dimensional change. This reduces mechanical stress but enhances sensitivity to potential-induced polarization and adsorption heterogeneity~\cite{he2023basics}. Together, these observations illustrate that SEI chemistry follows universal electrochemical rules—electrolyte breakdown, ion-migration-limited growth, and self-passivation—but device-specific boundary conditions (potential range, ion type, and surface reactivity) dictate its thickness, elasticity, and reversibility.

\begin{table}[htbp]
\centering
\caption{Comparison of SEI characteristics in batteries and supercapacitors.}
\begin{tabular}{p{3cm}p{5cm}p{5cm}}
\hline
\textbf{Feature} & \textbf{Battery SEI} & \textbf{Supercapacitor SEI-like Film} \\
\hline
Formation driver & Electrolyte reduction beyond stability limit & Local overpotential, defect-induced decomposition \\
Thickness & 10–100~nm & $<$10~nm \\
Stability & Quasi-static & Dynamic, reversible \\
Dominant composition & Inorganic (LiF, Li$_2$CO$_3$, Li$_2$O) & Mixed organic–inorganic (carbonates, nitriles, fluorides) \\
Primary function & Prevent further decomposition, enable ion intercalation & Modulate leakage, stabilize double layer \\
Mechanical behavior & Rigid, stress-sensitive & Quasi-elastic, self-healing \\
\hline
\end{tabular}
\end{table}

\subsection{Governing Parameters in Interphase Evolution}

The formation, composition, and stability of SEI layers depend on electrolyte formulation, operating voltage, temperature, and electrode surface characteristics.

\textbf{Electrolyte chemistry.}  
Solvent reduction potential, salt composition, and additives dictate SEI precursors. Low-viscosity solvents with wide electrochemical windows, such as acetonitrile, minimize decomposition but can yield thin SEIs of nitrile oligomers and anion fragments~\cite{chen2020improving}. Propylene carbonate–based electrolytes polymerize upon reduction, forming thicker elastic films. The choice of salt (TEABF$_4$, LiPF$_6$, NaTFSI) determines anion-derived inorganic species~\cite{tan2023structural}. Additives first optimized for Li-ion SEI stabilization, such as fluoroethylene carbonate (FEC) or vinylene carbonate (VC), are now used to regulate SEI growth in capacitors, underscoring transferable interfacial design principles~\cite{jiang2024engineering, tan2023structural}.

\textbf{Voltage window.}  
The applied potential defines the driving force for electrolyte decomposition. In batteries, operation intentionally exceeds the stability limit to form SEI; in capacitors, such excursions occur only locally. Below the limit, only EDL formation occurs; beyond it, SEI nucleation begins~\cite{gossage2023understanding, wang2022unraveling}. Overpotential exposure thickens the SEI and increases resistance. Optimizing voltage range balances energy density and durability, while surface heterogeneity modulates local reaction rates~\cite{ko2022electrode, ko2025degradation, chen2021overpotential}.

\textbf{Temperature.}  
Thermal effects influence both kinetics and stability. Elevated temperatures accelerate reactions but destabilize metastable organics, whereas low temperatures yield porous, resistive films~\cite{lee2016damage}. Long-term thermal cycling promotes compositional segregation where inorganic domains densify and polymeric regions soften~\cite{deshpande2017modeling, pavone2022reactivity}. Similar temperature-induced reconstructions occur in high-rate batteries, reinforcing the universality of interphase dynamics.

\textbf{Electrode surface states.}  
Surface chemistry and defect density determine initiation sites and adhesion strength. Graphitic carbons with basal planes form thin uniform SEIs; defective or functionalized carbons catalyze solvent reduction, yielding thicker, chemically diverse films~\cite{tan2023structural}. Transition-metal oxides and MXenes with variable terminations (–O, –F, –OH) actively participate in redox reactions that reshape SEI chemistry~\cite{zhang2024evolution}. Engineering surface functionality through doping or coatings thus provides a cross-platform strategy for controlling interphase morphology and stability~\cite{tan2023structural, pavone2022reactivity, barai2019impact}.

Overall, the physicochemical environment determines whether the SEI acts as a beneficial stabilizing layer or a resistive barrier. Integrating these parameters into unified electrochemical models will enable predictive interphase engineering across both batteries and supercapacitors. The following section (Section~3) examines the microscopic mechanisms that drive SEI nucleation, growth, and reconstruction, establishing the molecular basis for the interphase behaviors described above.

\section{Microscopic Mechanisms of SEI Formation and Evolution}

SEI formation is a universal electrochemical phenomenon arising whenever the applied potential drives an electrolyte beyond its thermodynamic stability window~\cite{adenusi2023lithium,suo2017solid,spotte2022towards}.
In both batteries and supercapacitors, the interphase originates from coupled electron-transfer, ion-migration, and molecular-reorganization events that spontaneously self-assemble at the electrode–electrolyte boundary~\cite{ding2024quantitative}.
The distinction lies in scale rather than mechanism: battery SEIs are thick and quasi-static, whereas capacitor SEIs are ultrathin, dynamic, and continuously reconstructed under cycling~\cite{wang2018review,an2016state}.
Understanding their shared microscopic origins is essential for cross-platform interphase engineering.

\subsection{Electron Transfer and Electrolyte Reduction Criteria}

SEI nucleation begins when the electrode Fermi level intersects the lowest unoccupied molecular orbital (LUMO) of the electrolyte, enabling electron tunneling into solvated molecules and initiating bond cleavage~\cite{li2023review,madhusudhanan2023revisiting}.  
The same criterion governs electrolyte reduction in both Li-ion batteries and carbon-based supercapacitors, although the effective overpotential and reaction timescales differ~\cite{ko2022electrode}.
Table~\ref{tab:SEIcomparison} summarizes the principal thermodynamic parallels.

\begin{table}[htbp]
\centering
\caption{Comparative features of SEI formation in batteries and supercapacitors.}
\label{tab:SEIcomparison}
\begin{tabular}{p{3cm}p{5cm}p{5cm}}
\hline
\textbf{Aspect} & \textbf{Batteries} & \textbf{Supercapacitors} \\
\hline
Driving potential & Electrolyte reduction at anode potentials ($<1$ V vs Li/Li$^+$) & Localized overpotential or defect-site fields ($1$–$3$ V window) \\
Film thickness & 10–100 nm, diffusion-limited growth & 1–10 nm, field-driven reconstruction \\
Dominant species & LiF, Li$_2$CO$_3$, Li$_2$O & Carbonates, nitriles, fluorides \\
Growth mode & Self-limiting, inorganic barrier & Dynamic, elastic dielectric \\
\hline
\end{tabular}
\end{table}

Electron injection into solvent or anion orbitals initiates decomposition of carbonate (EC, PC) or nitrile (ACN) molecules~\cite{wang2020reductive}.  
Regardless of electrolyte type, the resulting fragments—carbonates, alkoxides, nitrile oligomers—polymerize or condense on the surface, establishing the initial interphase~\cite{russell2016electrolyte,li2023review}.
Anion reduction (BF$_4^-$, PF$_6^-$, TFSI$^-$) produces inorganic fluorides and borates in both systems, yielding the dense, electronically insulating inner region that terminates further electron leakage~\cite{andersson2024initial}.  
Thus, SEI nucleation follows the same thermodynamic sequence in batteries and capacitors, differing only in potential range and material identity.

\subsection{Atomistic Decomposition and Fragment Condensation}

At the atomic scale, SEI formation proceeds through multiple, overlapping routes: solvent reduction, salt decomposition, and radical polymerization~\cite{banerjee2025formation}.  
In Li-ion cells, EC reduction forms Li$_2$CO$_3$ and polycarbonate fragments~\cite{weddle2023continuum,wang2001theoretical}; in TEABF$_4$/ACN capacitors, analogous pathways yield CN-based polymers and C–F moieties~\cite{kurzweil2008electrochemical}.  
Salt-derived anion cleavage (BF$_4^- \!\rightarrow\!$ LiF or C–F) provides inorganic nuclei that anchor the interphase.
These reactions are exothermic and locally heat the interface, promoting polymer growth much as observed during initial battery cycling~\cite{wang2025revisiting}.  
In ionic-liquid and water-in-salt electrolytes, coordinated water or cation fragments yield inorganic fluorides or hydroxides that emulate the protective SEIs of batteries~\cite{sun2021anion}.  
Hence, solvent and anion decomposition chemistry is fundamentally transferable between the two device classes.

\subsection{Heterogeneous Interphase Structure and Charge Selectivity}

Once decomposition products accumulate, they organize into a bilayered morphology: a dense, inorganic inner layer and a porous, polymeric outer matrix~\cite{yan2021nucleation}.  
In batteries, LiF/Li$_2$CO$_3$ layers dominate the inner region~\cite{wang2025revisiting}; in capacitors, C–F/BF$_x$O$_y$ and M–O–F domains play the same mechanical and electronic role~\cite{wu2025revealing}. 
The outer polymeric zone—composed of polycarbonate, nitrile, or ether species—acts as an ion-permeable, elastically compliant dielectric.  
This architecture self-regulates via potential-dependent current tunneling: when the film reaches a critical thickness $d$, the tunneling current density  
\[
J \!\propto\! \exp(-\beta d)
\]
drops below the decomposition threshold, halting further growth.  
Such self-passivation, long established for battery SEIs, equally governs the stabilization of capacitor interphases.

\subsection{Desolvation and Ion Migration Through Passivating Films}

Ion solvation governs both the onset of electrolyte reduction and the transport properties of the resulting SEI.  
At the interface, strong electric fields induce partial desolvation, exposing bare ions that readily accept or donate electrons~\cite{lee2025designing}.  
In batteries, Li$^+$ desolvation triggers EC reduction; in capacitors, partial desolvation of bulky cations such as tetraethylammonium yields solvent radicals that polymerize at the surface.
The degree of solvation controls the SEI composition: tightly bound solvates produce dense, inorganic-rich layers, while weakly solvated ions form thinner, more permeable films~\cite{lee2025designing,guo2024unifying,subramanian2005electroinitiated,li2024enhancing}.  
This relationship can be expressed by the correlation between solvation energy ($\Delta G_{\mathrm{solv}}$) and effective ionic conductivity ($\sigma_{\mathrm{ion}}$):
\[
\sigma_{\mathrm{ion}} \propto \exp\!\left(-\frac{|\Delta G_{\mathrm{solv}}|}{kT}\right),
\]
demonstrating that interfacial transport follows universal scaling across devices.  
Molecular-dynamics studies confirm similar desolvation barriers and coordination rearrangements at Li, Na, and carbon surfaces, reinforcing the mechanistic continuity.

\begin{figure}[htbp]
  \centering
  \includegraphics[width=0.9\textwidth,page=1]{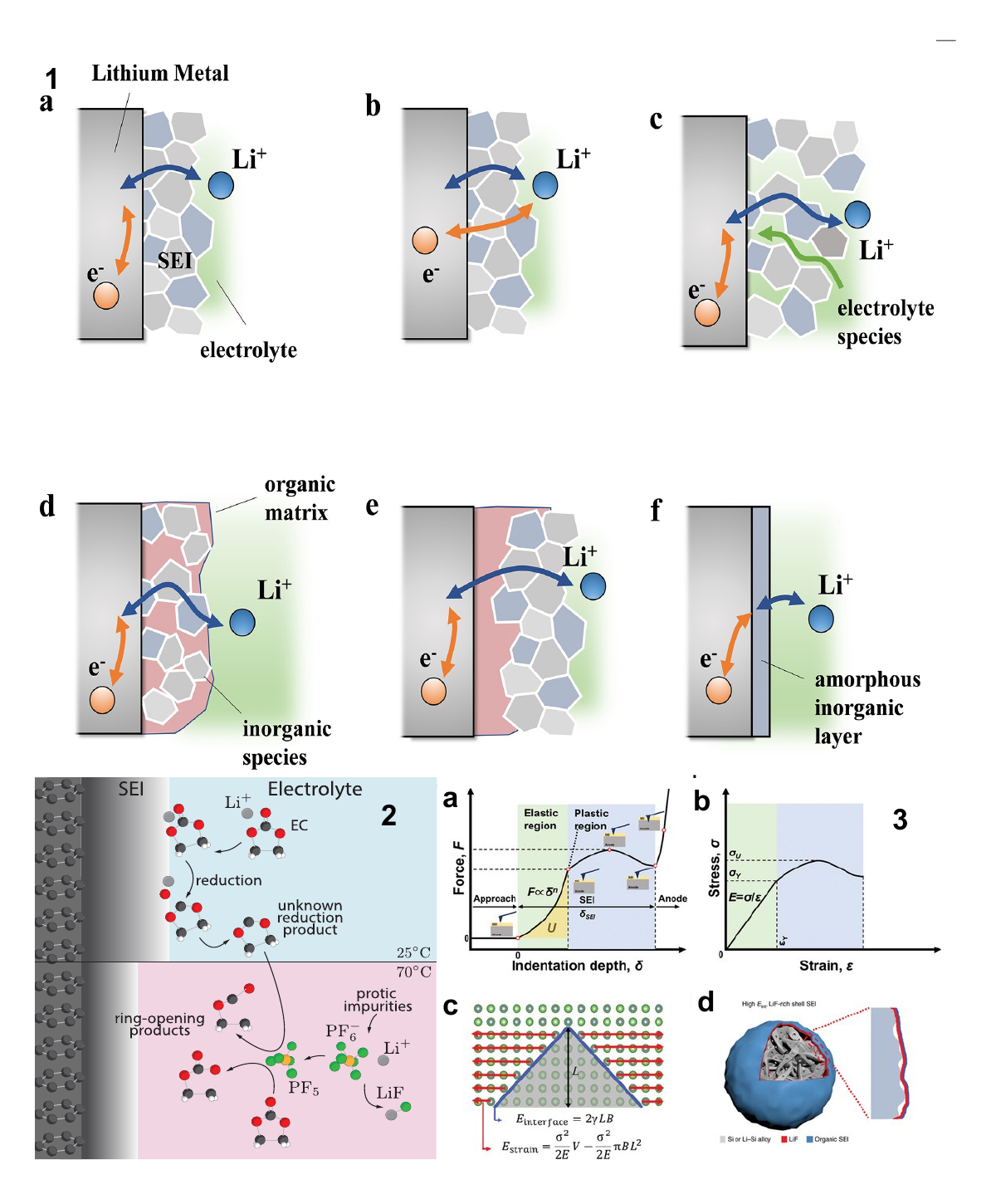}
\caption{Schematic overview of SEI formation across energy-storage systems, showing five common stages: (1) electron injection and inorganic nucleation\cite{nojabaee2021understanding}, (2) solvent/anion decomposition\cite{saqib2018decomposition}, and (3) mechanical relaxation\cite{li2025understanding}.}
  \label{fig:sec_3_1}
\end{figure}

\begin{figure}[htbp]
  \centering
  \includegraphics[width=0.9\textwidth,page=1]{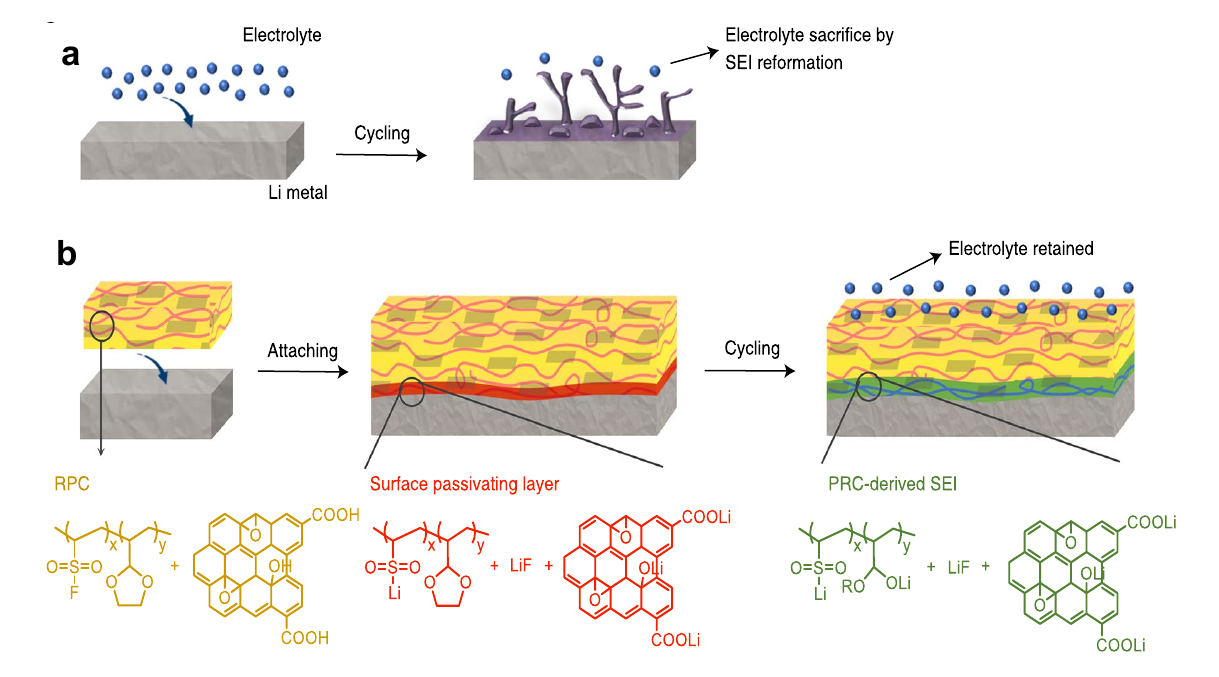}
\caption{Schematic overview of SEI formation across energy-storage systems via Polymer condensation\cite{gao2019polymer}.}
  \label{fig:sec_3_2}
\end{figure}

\subsection{Dynamic Reconstruction Under Electrochemical Stress}

The SEI continually evolves under electrochemical and thermal stimuli.  
Elevated temperature accelerates decomposition and diffusion, densifying the film in both systems~\cite{asl2010inspecting}; low temperature slows polymerization, yielding porous, resistive layers.  
During potential cycling, elastic compression and relaxation cause reversible thickness oscillations—observed by operando EQCM and cryo-TEM in both batteries and capacitors~\cite{kitz2018operando}. 
These oscillations follow diffusion–reaction kinetics similar to logarithmic growth laws~\cite{peak1972diffusion}:
\[
d(t) = k\,t^{1/2},
\]
where $k$ encapsulates ionic mobility and reaction rate constants.  
Such self-healing and reconstruction phenomena demonstrate that the SEI is not a static coating but a dynamic, viscoelastic interface adapting to electrochemical stress across all energy-storage systems.

\subsection{SEI-Mediated Modulation of Capacitance and Impedance}

The SEI critically influences electrochemical response by mediating ion accessibility and dielectric behavior.  
In both devices, its thickness and ionic conductivity determine the distance between ionic and electronic charge centers, thereby affecting differential capacitance and internal resistance~\cite{kulathuvayal2023ionic,chu2020advanced}. 
Nyquist plots exhibit a mid-frequency semicircle associated with SEI resistance ($R_{\mathrm{SEI}}$), whether measured in Li-ion cells or high-voltage capacitors.  
A thin, ion-permeable SEI minimizes leakage and stabilizes Coulombic efficiency, whereas a thick, resistive one increases relaxation times and lowers power density~\cite{morasch2024li,talian2024impedance}.

Spatial heterogeneity of the SEI also modifies local permittivity: inorganic-rich domains ($\varepsilon_r\!\approx\!5$–10) focus electric fields and promote ion layering, while polymeric regions ($\varepsilon_r\!\approx\!20$–40) smooth potential gradients.  
This heterogeneity generates two characteristic time constants in impedance spectra—a fast component from electric-double-layer charging and a slower one from ion migration through the SEI~\cite{sun2024enabling,chen2025heterogeneity}.  
In both batteries and capacitors, these coupled processes define the dynamic bottleneck for high-rate performance.

Furthermore, reversible redox rearrangements within the organic fraction of the SEI can contribute to pseudocapacitive charge storage.  
Such behavior, once noted in polymer-rich battery SEIs, equally appears in stabilized capacitor interfaces, where the SEI acts as an active, charge-mediating dielectric rather than a passive barrier~\cite{xu2022surface,hao2025structure}.

\subsection{Unified Framework for Interphase Formation}

Figure~\ref{fig:sec_3_1} \& \ref{fig:sec_3_2} schematically summarizes SEI formation across energy-storage systems, outlining five common stages:  
(1) electron injection, (2) solvent/anion decomposition, (3) inorganic nucleation, (4) polymeric condensation, and (5) mechanical relaxation~\cite{aoki2022effective,li2024understanding}.  
Differences between batteries and capacitors arise primarily from potential magnitude and cycling frequency, not from reaction identity.  


Across all electrochemical devices, the SEI embodies the same self-organization of matter under electric fields and chemical stress~\cite{an2016state,wu2023effect}.  
Recognizing this universality transforms the SEI from a device-specific artifact into a general design principle linking molecular chemistry to macroscopic performance.  
Having established the fundamental mechanisms of SEI formation, we now turn to the experimental methodologies that enable direct observation and characterization of these interphases in Section~4, followed by computational modeling frameworks in Sections~5 and~6 that predict and simulate SEI evolution across multiple scales.

\section{Experimental Characterization of the SEI}

Understanding the chemistry, morphology, and dynamics of the solid–electrolyte interphase (SEI) requires a combination of ex-situ structural analysis and in-situ or operando spectroscopy. 
Most techniques now used to probe SEI evolution were first developed in battery research—where stable interphases are intrinsic to operation—and have since been adapted to investigate the thinner, more dynamic SEI-like films in supercapacitors~\cite{capone2024operando,freitas2023combining,huang2019evolution}. 
Recognizing this shared experimental heritage enables a unified interpretation of interfacial chemistry across both systems and provides critical validation for atomistic simulations and continuum models discussed in Section~5~\cite{kitz2018operando,diddens2022modeling}.

\subsection{Ex-Situ Techniques for Interphase Characterization}

\textbf{X-ray Photoelectron Spectroscopy (XPS).}  
Originally established as the principal diagnostic for SEI analysis in Li-ion batteries, XPS remains the most widely used technique for probing elemental composition and oxidation states within interphases. Depth profiling by successive argon-ion sputtering reveals characteristic compositional gradients—LiF or Li$_2$CO$_3$-rich inner layers in batteries and inorganic fluorides or carbonates in capacitors\cite{shard2024practical, yakovlev2005interfacial}. In carbon-based supercapacitors, XPS detects C–O, C=O, and O–C=O functionalities arising from solvent oxidation, while F 1s and B 1s peaks confirm BF$_4^-$ or PF$_6^-$ decomposition\cite{andersson2002surface}. Quantitative peak-shift analysis further provides information on local potential drops and the electronic insulation conferred by the SEI layer. Thus, XPS offers a common compositional framework for comparing SEI chemistry across batteries and capacitors.

\textbf{Fourier-Transform Infrared Spectroscopy (FTIR).}  
FTIR identifies vibrational fingerprints of organic components. Bands at $\sim$1700 cm$^{-1}$ (C=O), $\sim$1100 cm$^{-1}$ (C–O–C), and carbonate stretches signal solvent decomposition and polymerization~\cite{shard2024practical}. In battery studies, these signatures verified polymeric carbonates and alkoxides; analogous features in supercapacitors indicate nitrile- or carbonate-based oligomers formed from acetonitrile or propylene carbonate breakdown. Mapping FTIR intensity across electrodes reveals heterogeneity linked to potential gradients and pore accessibility\cite{chang2011effect, ridier2016enhanced}.

\textbf{Raman Spectroscopy.}  
Raman scattering simultaneously probes electrode disorder and interphase evolution. The D/G-band ratio (I$_D$/I$_G$) reflects graphitic defect density, while new bands between 1200–1500 cm$^{-1}$ correspond to polymeric species within the SEI\cite{an2016state}. Techniques such as surface-enhanced Raman spectroscopy (SERS) and tip-enhanced Raman (TERS), originally applied to map SEI growth on graphite anodes, now resolve nm-scale chemistry of dynamic SEI films in capacitors\cite{schultz2014tip, schmid2013nanoscale}.

\textbf{Nuclear Magnetic Resonance (NMR).}  
Solid-state $^{13}$C, $^{19}$F, and $^1$H NMR spectroscopy provides direct insight into molecular bonding and dynamics. Broad peaks from immobilized polymeric fragments contrast with sharp resonances from free electrolyte molecules, allowing estimation of cross-linking and segmental mobility\cite{wang2020reductive}. Approaches such as dynamic nuclear polarization (DNP), initially implemented to enhance SEI sensitivity in Li-ion batteries, have begun to uncover weak interfacial polymer–electrolyte coupling in supercapacitors.

\textbf{Time-of-Flight Secondary Ion Mass Spectrometry (ToF-SIMS).}  
ToF-SIMS delivers nm-scale surface sensitivity and isotopic resolution, detecting fragment ions that pinpoint SEI precursors\cite{an2016state}. In both systems, cryogenic operation mitigates beam damage and preserves fragile species\cite{vilkevivcius2024tuning}. Depth-resolved mapping visualizes heterogeneity—from LiF mosaics in batteries to mixed organic–fluoride domains in capacitors—providing critical benchmarks for validating reactive-MD predictions of interphase composition.

\subsection{In-Situ Monitoring of SEI Formation and Evolution}

\textbf{Electrochemical Quartz Crystal Microbalance (EQCM).}  
EQCM, developed to track mass changes during battery SEI growth, now measures real-time ion adsorption and SEI formation in capacitors with nanogram sensitivity\cite{qin2020eqcm}. Monotonic mass increases indicate film deposition, while oscillatory trends correspond to reversible ion intercalation. Combining EQCM with electrochemical impedance spectroscopy (EIS) distinguishes elastic mass loading from viscoelastic film formation, unifying the interpretation of interphase kinetics.

\textbf{Electrochemical Impedance Spectroscopy (EIS).}  
EIS quantifies charge-transfer resistance and ion diffusion within SEI layers. The appearance of an additional semicircle at intermediate frequencies—associated with SEI resistance ($R_{\mathrm{SEI}}$)—was first observed in battery electrodes and is now routinely analyzed in capacitors\cite{tan2023structural}. Monitoring EIS evolution during cycling reveals how interphase maturation parallels the transition from thick, quasi-static SEIs in batteries to thin, adaptive layers in capacitors.

\textbf{In-situ Raman and Infrared Spectroscopy.}  
Operando Raman and attenuated-total-reflectance (ATR) infrared spectroscopy capture real-time bond formation and cleavage under bias. Frequency shifts in C$\equiv$N or S=O stretching modes trace reversible ion solvation or irreversible decomposition leading to SEI formation\cite{zhou2023unraveling}. These techniques, originally applied to monitor electrolyte degradation in Li-ion cells, now enable spatially resolved mapping of interfacial heterogeneity in capacitors.

\textbf{Cryogenic Transmission Electron Microscopy (cryo-TEM).}  
Cryo-TEM, which revolutionized SEI imaging in batteries, preserves native interfacial structures by rapid freezing\cite{zhang2023revealing}. Its adaptation to supercapacitors has revealed ultrathin, self-healing SEIs in ionic-liquid systems, providing direct evidence that reversible reconstruction is a universal property of electrochemical interphases.

\textbf{Atomic Force Microscopy (AFM) and Electrochemical AFM (EC-AFM).}  
AFM quantifies topography and mechanical stiffness of SEIs with nanometer precision. Force–distance curves and current-sensing AFM (CS-AFM) map film elasticity and local conductivity\cite{tan2023structural, izawa2018crystallization}. EC-AFM, conducted under potential control, dynamically visualizes SEI evolution in liquid environments—linking morphology, adhesion, and charge-transfer resistance across both device types.

\begin{figure}[htbp]
  \centering
  \includegraphics[width=0.95\textwidth,page=1]{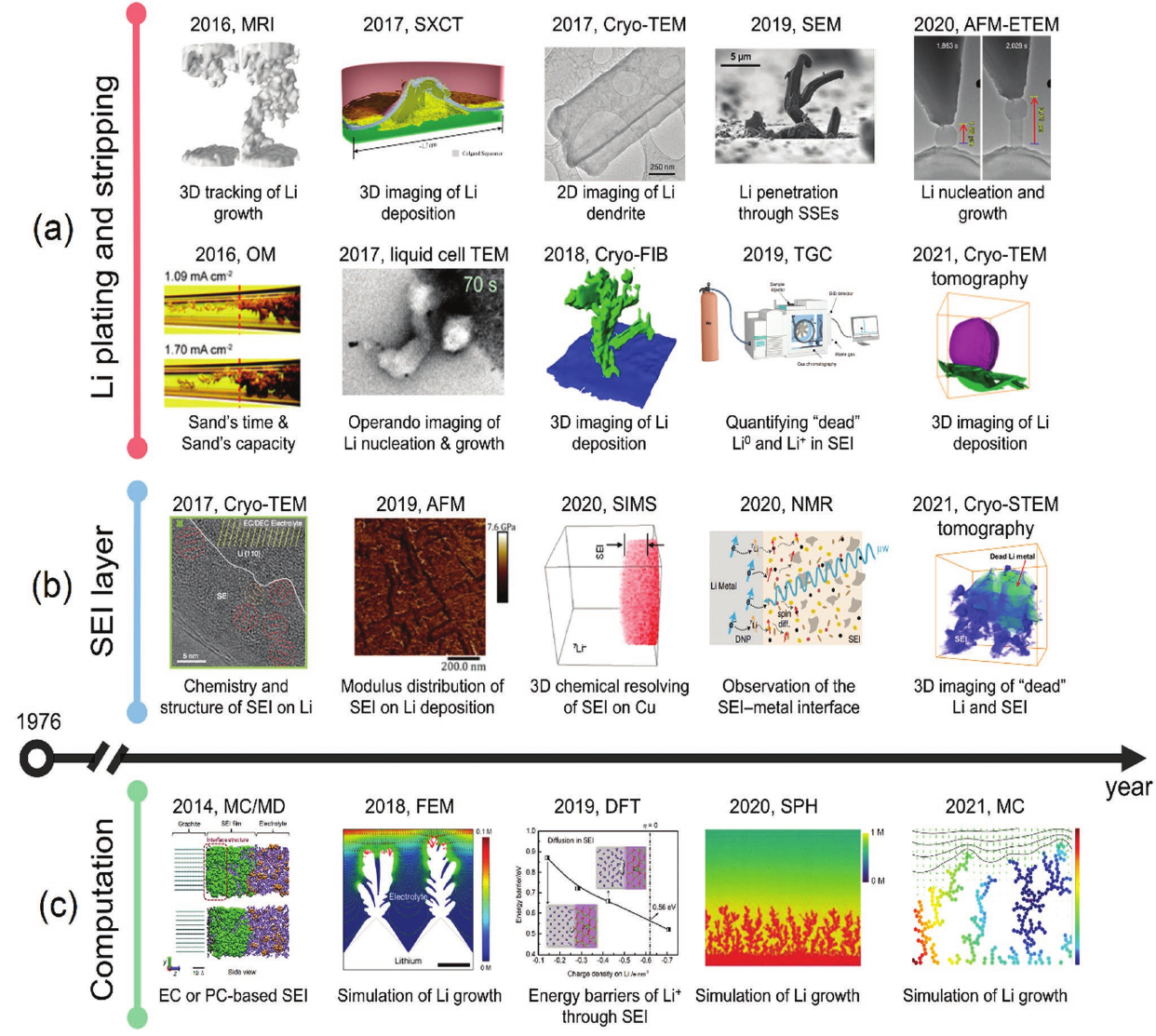}
\caption{ Recent advances in mechanistic understandings of Li deposition and SEI using advanced analytical techniques and theoretical modelling 
methods. a) Advanced characterization techniques for analyzing Li plating and stripping behaviors include Magnetic resonance imaging (MRI), SXCT), Cryo-TEM, SEM, AFM-ETEM, OM, liquid-cell TEM, Cryo-FIB, TGC, and Cryo-TEM. b) Analysis of SEI involves cryo-TEM, AFM, SIMS, NMR, and cryo-STEM tomography. c) Theoretical modelling methods include MC/MD, finite element method 
(FEM), DFT(B), SPH, and MC for calculation of the diffusion energy barriers and simulation of the Li deposition behavior and SEI structure\cite{xu2022promoting}.}
  \label{fig:charac_sec_4}
\end{figure}

\subsection{Cross-Technique Validation and Data Synthesis}

Correlating chemical and electrochemical data is crucial for understanding SEI function in any electrochemical system. In both batteries and capacitors, combining XPS-derived composition with EIS-extracted $R_{\mathrm{SEI}}$ values links specific motifs (fluorides, carbonates) to ionic resistivity\cite{tan2023structural}. FTIR and Raman features associated with polymeric carbonates often coincide with higher leakage currents or reduced capacitance, indicating partial electronic conductivity through conjugated species. EQCM mass-change profiles synchronized with cyclic-voltammetry (CV) peaks directly reveal the competition between capacitive ion adsorption and interphase growth\cite{chen2024design}. These cross-correlations provide quantitative descriptors—such as “interphase efficiency” or “ionic utilization ratio”—that unify SEI analysis across technologies.

Multimodal platforms now integrate spectroscopy, microscopy, and electrochemistry in real time. Operando Raman–EIS coupling simultaneously tracks vibrational and impedance signatures, while synchrotron-based soft X-ray absorption spectroscopy (sXAS) resolves element-specific oxidation states\cite{fan2025operando}. These hybrid systems establish feedback loops between experiment and simulation, enabling parameter calibration for molecular-dynamics and DFT models and strengthening the link between mechanistic insight and performance metrics.

\subsection{Comparative Studies Across Electrode and Electrolyte Systems}

On \textbf{carbonaceous electrodes}, XPS and EQCM studies in both battery and capacitor contexts reveal oxygen- and fluorine-containing SEIs forming above 2.5 V in organic electrolytes such as TEABF$_4$/ACN\cite{shen2015situ}. The interphase suppresses solvent oxidation but raises series resistance. In aqueous KOH or Na$_2$SO$_4$, Raman and AFM detect transient hydroxide layers that dissolve upon potential reversal—demonstrating the reversible nature of aqueous SEIs\cite{zhang2020situ}.

For \textbf{transition-metal-oxide electrodes} (e.g., MnO$_2$, NiO, V$_2$O$_5$), XPS and FTIR reveal hybrid SEIs combining metal–oxygen species and organic fragments\cite{liang2024situ}. Such interphases facilitate proton or cation insertion in both batteries and capacitors, though excessive growth hinders diffusion and causes capacitance fading.

\textbf{Ionic-liquid-based systems} exhibit inorganic-rich SEIs dominated by F- and S-containing fragments (TFSI$^-$ or PF$_6^-$ derivatives)\cite{jankowski2019functional}. Cryo-TEM and ToF-SIMS confirm dense coverage that stabilizes both Li-metal and carbon interfaces. In solid or gel polymer electrolytes, FTIR and NMR reveal polymer–electrode cross-linking and hydrogen-bond networks controlling interfacial ion mobility\cite{capone2024operando, guo2021reactivity}. These case studies underscore that SEI chemistry reflects universal electrochemical principles—modified by electrode material, electrolyte chemistry, and potential range—rather than device category.

\subsection{Sources of Uncertainty and Inter-Laboratory Variability}

Despite progress, SEI characterization faces common limitations in both research domains. Ultrathin, reactive films challenge sample preparation and air-free transfer\cite{mohanty2023iron}. Ex-situ methods capture static snapshots, whereas in-situ techniques often trade spatial resolution for time resolution\cite{guoguang2019situ}. Discrepancies among laboratories arise from differences in cell geometry, electrolyte purity, and reference calibration\cite{rui_2023}. The adoption of glovebox-integrated, cryogenic, and statistically replicated protocols now improves reproducibility. Moreover, coupling experimental outputs with simulation benchmarks—such as radial distribution functions and predicted binding energies—provides cross-validation and minimizes interpretational ambiguity.

\begin{table}[htbp]
\centering
\caption{Representative techniques for SEI characterization and their cross-system adaptation.}
\begin{tabular}{p{3cm}p{4cm}p{4cm}p{4cm}}
\hline
\textbf{Technique} & \textbf{Battery Application} & \textbf{Supercapacitor Adaptation} & \textbf{Insight Bridged} \\
\hline
XPS & Depth profiling of LiF/Li$_2$CO$_3$ SEI & Detection of fluorinated/nitrile films & Composition and oxidation states \\
FTIR/Raman & Organic carbonate identification & Nitrile/carbonate vibrations & Functional-group evolution \\
Cryo-TEM & SEI morphology on Li metal & Dynamic reconstruction in IL-based capacitors & Structural reversibility \\
EQCM/EIS & Mass and impedance evolution & Film growth vs ion adsorption & Kinetic coupling \\
AFM/EC-AFM & Nanomechanical mapping & Interphase elasticity and adhesion & Mechanical response \\
\hline
\end{tabular}
\end{table}

\subsection{Summary}

The migration of SEI-characterization methods from batteries to supercapacitors has unified interphase science across energy-storage technologies. 
Techniques once reserved for thick, inorganic-rich films now resolve nanometer-thin, reversible SEIs in capacitors. 
Establishing standardized cross-system protocols—consistent potential referencing, cryogenic handling, and quantitative multimodal analysis—will enable reproducible comparison and accelerate the integration of experimental data with computational modeling.

The experimental insights presented in this section provide the empirical foundation for the multiscale modeling frameworks discussed in Section~5, where atomistic simulations, continuum models, and hierarchical coupling strategies translate observed interphase properties into predictive design tools.
Section~6 then explores how machine learning amplifies these modeling capabilities, enabling data-driven SEI optimization.
Finally, Section~7 synthesizes the materials and interface engineering strategies that translate experimental and computational insights into practical device improvements.

\section{Materials and Interface Engineering Strategies}

Performance and durability in both batteries and supercapacitors are governed by the interplay among electrodes, electrolytes, separators, and fabrication protocols.  
Although their energy-storage mechanisms differ, the fundamental challenge is the same: controlling interfacial reactions to form a stable, ion-conductive, and electronically insulating solid–electrolyte interphase (SEI)~\cite{zhou2020real,zhang2024single}.  
This section consolidates materials selection, electrolyte design, surface modification, artificial interphase strategies, and mechanical stabilization approaches that have traditionally evolved in battery science but are now being systematically translated to supercapacitors, establishing a coherent framework for cross-platform SEI engineering.

The organization follows a logical progression: electrode materials and their surface engineering (Section~7.1), electrolyte chemistry and additive design (Section~7.2), artificial interphase approaches (Section~7.3), mechanical and thermal stabilization (Section~7.4), and fabrication architectures (Section~7.5).
Each subsection highlights how insights from batteries inform capacitor design, and vice versa.

\subsection{Active Material and Interface Modification Approaches}

The electrode acts as both the electronic conductor and the reactive template on which electrolyte decomposition initiates.  
In batteries, electrode surfaces catalyze reduction of solvent molecules and salts, forming SEI layers that regulate ion transport.  
A similar, though thinner and more dynamic, interphase develops in supercapacitors under high potentials~\cite{yang2023solid,hess2025surface}.  
Understanding this shared interfacial chemistry requires analyzing structure, composition, and surface functionality of the electrode, as well as deliberate surface modification strategies.

\subsubsection{Carbon-Based Electrodes}

Carbon materials remain the foundation of both Li-ion battery anodes and electric double‐layer capacitors (EDLCs).  
Activated carbon, carbon nanotubes (CNTs), graphene, and carbide‐derived carbons provide large surface areas and excellent conductivity~\cite{sk2023green,kraiwattanawong2022review}.  
In batteries, defects and oxygenated groups nucleate SEI formation; in capacitors, analogous surface terminations guide adsorption and mild solvent decomposition, yielding ultrathin SEI films~\cite{kado2019advanced}.  
Thus, tuning defect density and heteroatom doping (N, S, P) emerges as a universal handle for modulating SEI chemistry, balancing capacitance retention with interfacial stability~\cite{wang2025minimizing, li2024structural}.

\subsubsection{Metal Oxide and Conducting Polymer Systems}

Transition‐metal oxides such as MnO$_2$, NiO, and V$_2$O$_5$—and their analogs in battery cathodes—share redox‐active surfaces that catalyze both beneficial charge transfer and parasitic electrolyte decomposition~\cite{shaheen2024recent, pan2023application}.  
The resulting hybrid SEIs contain inorganic (M–O, M–F) and polymeric components that regulate ion transport~\cite{marr2014electrochemical, tseng2011tuning}.  
Stabilization strategies pioneered in batteries—surface coating with carbon or ALD oxides, and defect moderation—are now applied to pseudocapacitive electrodes to achieve similar durability and SEI uniformity~\cite{tan2023structural, wan2024designing}.

\subsubsection{Nanostructured Composites}

Composite electrodes such as graphene–MnO$_2$ and CNT–PPy integrate conductivity and redox activity~\cite{yu2024recent, huang2018synthesis}.  
Their multiphase interfaces resemble those in composite battery electrodes, where heterogeneous electronic environments induce spatially varying SEI composition~\cite{bedrov2020multiscale, wang2025electric}.  
Managing these local variations—through interface functionalization or graded heterostructures—has become a cross-cutting strategy for both technologies.

\subsubsection{Emerging Electrode Families}

MXenes, metal–organic frameworks (MOFs), and covalent–organic frameworks (COFs) are bridging materials that blur the line between capacitor and battery electrodes~\cite{chen2025compositing}.  
MXenes exhibit metallic conductivity and surface terminations (–O, –OH, –F) that facilitate controlled SEI formation in both Li-ion and solid-state supercapacitors~\cite{naren2024fluorinated}.  
MOF- and COF-derived carbons offer tunable porosity and heteroatom content, improving charge storage and interfacial uniformity~\cite{wang2025minimizing}.  
Their adaptability highlights a universal principle: surface chemistry, not device classification, governs SEI behavior.

\subsubsection{Surface Chemistry Engineering}

Functional-group engineering provides an atomistic handle on interphase chemistry.  
Oxygenated carbons (C–O, C=O) improve wettability and ion accessibility but can catalyze side reactions if uncontrolled; nitrogen or fluorine doping, by contrast, lowers surface free energy, anchors inorganic fragments, and suppresses solvent decomposition~\cite{singh2024graphene,wang2025structural}. 
In both batteries and capacitors, heteroatom doping modifies the electronic density of states near the Fermi level, shifting the onset potential for electrolyte reduction and enabling more benign, self-passivating SEI evolution.

Conformal coatings—such as atomic-layer-deposited (ALD) oxides, molecular-layer-deposited (MLD) polymers, polydopamine (PDA), or graphene-oxide (GO) films—pre-form artificial buffer layers that moderate the first contact between electrode and electrolyte~\cite{tan2023structural,liu2025surface}.
Thin ($<$5 nm) alumina or titania ALD coatings, proven to stabilize Li metal and Si anodes, have direct analogues in supercapacitors where they homogenize the interfacial field, prevent localized ion crowding, and mechanically reinforce porous structures~\cite{zheng2025coupled}.  
Recent advances in nanoscale electrode architecture—hierarchical porosity, vertically aligned nanotubes, and core–shell heterostructures—enable simultaneous control of ion transport and SEI uniformity.

Surface topology and curvature also influence SEI homogeneity.  
Hierarchically porous or scaffolded electrodes distribute ionic flux uniformly, reducing hot spots that trigger uncontrolled film growth~\cite{liu2025surface}.
For flexible or solid-state configurations, the combination of engineered surface morphology and compliant gel electrolytes produces solid–solid interfaces with enhanced mechanical tolerance~\cite{zhang2024single}.

\subsubsection{Structure–Interphase Relationships}

Across both systems, the electrode's microstructure and surface modification dictate the morphology and chemistry of the SEI.  
Planar graphitic surfaces favor uniform organic SEIs; rough or catalytically active oxides promote thicker, inorganic-rich films that raise resistance~\cite{tan2023structural}.  
Pore curvature, confinement, and doping tune local electric fields and electron density, leading to more homogeneous passivation layers.  
A mechanically stable and chemically uniform SEI ensures efficient ion transport and long-term cycling, whereas heterogeneity accelerates degradation.  
Thus, interfacial design principles—defect control, doping, surface functionalization, and conformal coatings—transcend device categories and form a universal toolkit for SEI engineering.

\subsection{Electrolyte Formulation for Controlled Interphase Formation}

Electrolyte formulation is the second axis of SEI control.  
The electrolyte acts as both reactant and architect of the SEI, supplying the molecular precursors from which the interphase self-assembles~\cite{ko2022electrode,jamil2024metal}.
Regardless of whether the system stores charge through ion adsorption or Faradaic reactions, electrolyte decomposition initiates interphase formation.  
Concepts first proven in battery electrolytes are now being systematically adapted to capacitors, revealing consistent mechanisms governing SEI chemistry.

\subsubsection{Role of Electrolytes in Interphase Formation}

The electrolyte conducts ions while insulating electrons between electrodes.  
Its dielectric constant, ion‐solvation strength, and electrochemical stability window (ESW) define both energy density and SEI composition~\cite{azmi2023electrochemical, krishnan2022effect}.  
In both devices, wide-ESW electrolytes (organic or ionic‐liquid) permit higher voltages but trigger reductive or oxidative decomposition that yields passivating films~\cite{zhang2022elucidating}.  
Therefore, electrolyte chemistry simultaneously enables performance and dictates interphase evolution.

\subsubsection{Solvent and Salt Selection Criteria}

Aqueous electrolytes (Na$_2$SO$_4$, KOH) offer high conductivity but narrow voltage windows, forming transient hydroxide SEIs~\cite{pathak2024high, zhu2022usefulness}.  
“Water‐in‐salt” formulations extend this window and generate inorganic‐rich SEIs analogous to those in aqueous batteries~\cite{chai2023achieving}.  
Organic solvents such as acetonitrile (ACN) or propylene carbonate (PC) paired with TEABF$_4$ yield higher voltages but undergo anion‐driven decomposition forming fluorinated SEIs similar to LiPF$_6$-based systems~\cite{zhang2016insights}.  
Room‐temperature ionic liquids (RTILs) further stabilize interfaces through intrinsic ion pairing and preorganized solvation shells~\cite{sayah2022super}.  
These parallels indicate that solvent chemistry universally governs SEI structure, independent of device label.

\subsubsection{Film-Forming Additives}

Additives such as vinylene carbonate (VC) and fluoroethylene carbonate (FEC), long established in Li-ion batteries, now serve as interphase stabilizers in supercapacitors~\cite{yohannes2019sei}.  
They promote early formation of uniform, fluorinated SEIs that suppress continuous electrolyte degradation.  
Similarly, corrosion inhibitors or pH regulators in aqueous systems mirror redox‐mediated additives in batteries~\cite{wang2023regulating}.  
Machine‐learning‐guided screening of additive–solvent–electrode combinations is emerging as a shared optimization tool across both research domains~\cite{rahmanian2025electrolyte}.  

\subsubsection{Electrolyte Structure–SEI Morphology Relationships}

Highly solvated ions favor polymeric, organic‐rich SEIs; desolvated ions yield dense, inorganic layers~\cite{quan2021unveiling}.  
This solvation–desolvation balance—previously quantified for Li$^+$ and Na$^+$ systems—equally determines the film composition on carbon electrodes in capacitors.  
Such universal scaling between ion solvation energy and SEI density provides a bridge for modeling interphase formation across energy‐storage technologies.

\subsubsection{Electrolyte–Separator Coupling}

Electrolyte and separator engineering collectively define ionic pathways and interfacial stability.  
Polyolefin (PP/PE) and PVDF separators ensure mechanical strength, while oxide‐ceramic coatings (Al$_2$O$_3$, TiO$_2$) or cellulose membranes enhance wettability and dielectric robustness~\cite{kong2022metal, li2023cellulose}.  
These design rules—originally optimized for lithium batteries—are now applied to supercapacitors to mitigate leakage and enable safe high-voltage operation.

\subsection{Architectural Design Principles}

Device architecture transforms material‐level control into macroscopic performance.  
Symmetric cells with identical electrodes mirror the balanced configurations of dual-ion batteries, whereas asymmetric and hybrid designs emulate full cells by coupling capacitive and Faradaic electrodes~\cite{shao2018design}.  
The same interfacial concerns—contact resistance, electrolyte wetting, and mechanical adhesion—govern both technologies.  
Optimization therefore relies on common engineering metrics: minimized IR drop, uniform current distribution, and stable SEI adhesion.

\subsection{Manufacturing and Structure–Property Relationships}

Electrode fabrication directly impacts SEI nucleation and evolution.  
Active materials (carbon, oxide, or polymer) are typically mixed with conductive additives (carbon black, CNTs) and binders such as PVDF or PTFE~\cite{simon2008materials, wang2012review}.  
The slurry is cast onto metallic current collectors (Al, Ni, stainless steel) and processed by doctor‐blade or drop‐casting, then dried and compressed for uniform thickness.  
Binder‐free architectures—graphene foams, carbon cloths, vertically aligned CNT arrays—minimize inactive mass and enhance conductivity~\cite{zhang2025review}.  
These structural motifs, analogous to 3D current collectors in batteries, promote homogeneous SEI formation and mechanical stability.

\subsubsection{Characterization and Cross-System Correlation}

Structural and chemical analyses connect fabrication parameters with interfacial properties.  
Microscopy (SEM/TEM) visualizes pore architecture and SEI coverage; BET quantifies accessible area; XRD and Raman assess crystallinity and defect density; XPS and FTIR identify chemical species~\cite{lefebvre2019experimental, matuana2001surface}.  
These same diagnostics—standard in both battery and capacitor research—provide a unified dataset for correlating morphology with electrochemical behavior.  
Integrating such multimodal data through statistical or machine‐learning models establishes transferable descriptors (e.g., SEI thickness, roughness, and impedance) that enable predictive interphase design.

\subsection{Summary}

Material and interface engineering strategies that stabilize SEIs in batteries—controlled surface reactivity, electrolyte additive tuning, nanostructured architecture, artificial coatings, and mechanical stabilization—are directly translatable to supercapacitors.  
The convergence of electrode and electrolyte design principles across these technologies demonstrates that SEI formation follows universal chemical and mechanical rules rather than device-specific phenomena.  
By integrating experimental workflows (Section~4) and computational models (Sections~5 and~6), researchers can derive cross-platform descriptors linking composition, morphology, and transport.  
These descriptors provide the foundation for connecting SEI microstructure to macroscopic device performance, as discussed in Section~8.

\section{Multiscale Modeling and Simulation of SEI Formation}

Understanding the solid–electrolyte interphase (SEI) requires bridging phenomena that span more than ten orders of magnitude in time and length.  
Bond cleavage and charge transfer occur at the quantum level, while film growth and mechanical evolution unfold across nanometers to micrometers and over thousands of cycles.  
Modeling frameworks must therefore integrate electronic-structure theory, molecular dynamics (MD), continuum transport, and device-scale formalisms within a coherent hierarchy~\cite{bedrov2020multiscale,choi2016atomistic}.  
Although the materials differ—Li metal anodes in batteries versus carbon electrodes in supercapacitors—the governing processes of electron tunneling, ion migration, and interphase reconstruction follow remarkably similar multiscale couplings~\cite{zhang2024liquid,wang2024mechanical}.
This section reviews computational approaches that link the microscopic mechanisms described in Section~3 with the experimental observables from Section~4, establishing predictive frameworks for SEI design.  

\subsection{Atomistic Simulation Methods}

\begin{figure}[htbp]
  \centering
  \includegraphics[width=1\textwidth,page=1]{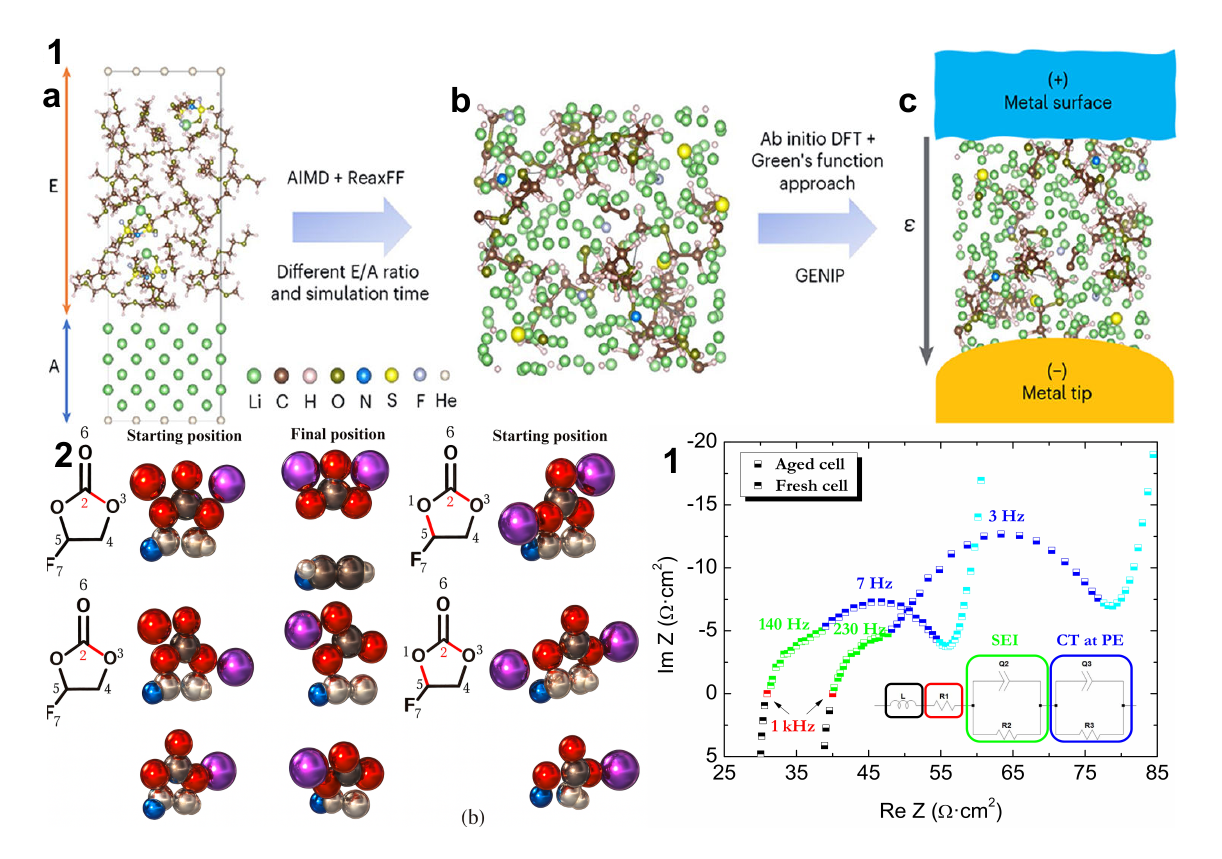}
\caption{(1) Atomistic modeling workflow: (a) AIMD/ReaxFF simulations capturing electrolyte decomposition under varying electron-to-atom (E/A) ratios; (b) structural evolution of SEI precursor species; and (c) GENIP-based ab initio DFT + Green’s-function framework used to evaluate interfacial electronic properties between the SEI and metal surface\cite{xu2023direct}.
(2) The first type of initial bond breaking mechanism of FEC leading to the formation of SEI\cite{zhang2020engineering}.
(3) Electrochemical impedance spectra (EIS) of fresh vs. aged cells, highlighting characteristic frequency regimes associated with SEI resistance and charge transfer processes at the porous electrode (CT at PE)\cite{ovejas2018impedance}.}
  \label{fig:sec_6_1}
\end{figure}

MD provides the fundamental atomistic perspective on SEI structure and dynamics.  
Atoms are treated as classical particles whose trajectories obey Newton’s equations, evaluated through empirical or reactive interatomic potentials.\cite{zeng2023molecular,yu2023constant,magnussen2019toward}  
In both batteries and supercapacitors, MD captures ion adsorption, desolvation, and transport near polarized surfaces, thereby resolving the molecular origins of electric‐double‐layer formation and early electrolyte decomposition.\cite{he2025bio,pean2015confinement,tan2024decoding}  
Analyses of velocity–autocorrelation and residence‐time functions yield diffusion coefficients and relaxation constants that can be directly compared with electrochemical impedance or cyclic‐voltammetry measurements.  
These correlations position MD as the molecular bridge linking interfacial structure with experimentally measurable capacitance and resistance.\cite{singer2017molecular} The integrated atomistic–electronic workflow connecting AIMD/ReaxFF decomposition pathways to experimental impedance signatures is summarized in Fig.~\ref{fig:sec_6_1}.

\subsection{Electrostatic Treatment of Polarized Interfaces}

Capturing realistic electrode polarization is central to simulating SEI nucleation.  
Constant‐charge (CCM) and constant‐potential (CPM) models remain the two canonical approaches.  
CCM assigns fixed atomic charges and thus reproduces qualitative ion layering but underestimates capacitance under strong electric fields.  
CPM dynamically adjusts atomic charges to maintain an imposed electrode potential, reproducing metallic screening and local charge redistribution.\cite{yang2017reliability,zeng2024constant}  
Recent extensions—generalized (GCPM), heteroatomic (HCPM), and moment‐tensor (mCPM) formulations—incorporate DFT‐derived electronegativity and hardness, enabling accurate simulation of doped carbons, MXenes, and Li metal.\cite{chaudhari2025machine,li2025enabling}  
Such self‐consistent polarization models now delineate the spatial onset of electron tunneling and electrolyte reduction, providing mechanistic routes to SEI nucleation in both capacitors and batteries.\cite{li2025enabling}

\subsection{Chemical Reactivity in Force-Field Frameworks}

\begin{figure}[htbp]
  \centering
  \includegraphics[width=1\textwidth,page=1]{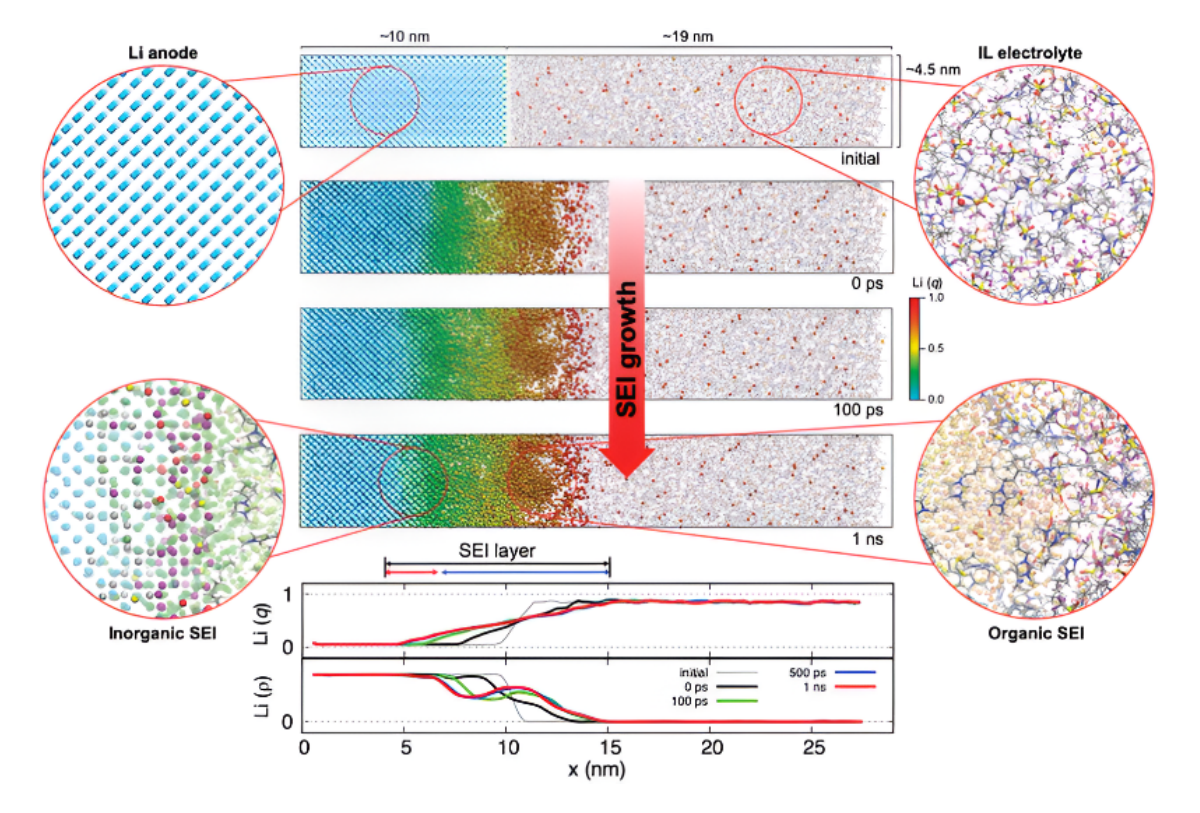}
\caption{Schematic overview of The formation of the SEI layer from the ReaxFF MD simulation at 300K The initial system\cite{yang2023characterization}.}
  \label{fig:sec_6_2}
\end{figure}

The reliability of any MD study ultimately rests on its interatomic potential.  
Fixed‐charge force fields such as OPLS, AMBER, and COMPASS are computationally efficient but non‐reactive.  
Polarizable models, including induced‐dipole and Drude‐oscillator schemes, introduce electronic response, while charge‐equilibration (QEq) and fluctuating‐charge (FQ) approaches extend this to metallic polarization.\cite{dai2010enthalpies,riniker2018fixed}  
Reactive potentials such as ReaxFF explicitly describe bond formation and scission, thereby allowing simulation of electrolyte decomposition and SEI film growth.\cite{han2016development}  
In Li‐ion systems, ReaxFF has clarified ethylene carbonate reduction to Li$_2$CO$_3$, whereas in supercapacitors it reproduces acetonitrile and ionic‐liquid fragmentation that yields CN‐ or F‐containing interphases.\cite{wang2020reductive,hossain2020lithium}  
These studies consistently reveal heterogeneous nucleation—initial anchoring of inorganic fragments on the electrode followed by polymeric condensation. A representative ReaxFF simulation of SEI nucleation and early film formation is illustrated in Fig.~\ref{fig:sec_6_2}, showing the molecular evolution captured during electrolyte decomposition.

Machine‐learning (ML) potentials such as DeepMD, NequIP, and SpookyNet offer near‐\textit{ab initio} accuracy with classical scalability.\cite{unke2021spookynet}  
Trained on DFT trajectories for representative battery (LiPF$_6$/EC) and capacitor (TEABF$_4$/ACN) chemistries, these models reproduce many‐body polarization, reactive events, and charge redistribution across diverse interfaces.  
Integration with active‐learning and coarse‐graining protocols has extended the accessible time window from nanoseconds to milliseconds, allowing direct simulation of interphase evolution under operational conditions.\cite{wilson2023anisotropic}  

\subsection{Mesoscale and Macroscopic Film Evolution Models}

At larger scales, continuum and kinetic models treat the SEI as an effective mixed ionic–electronic conductor.  
Classical battery formulations by Peled and Newman describe diffusion‐limited thickening with $d \!\propto\! t^{1/2}$,\cite{vishweswariah2025evaluation} while supercapacitor adaptations incorporate the rapid charge–discharge cycles and mechanical elasticity of thin interphases.  
The temporal evolution of film thickness can be expressed as
\[
\frac{\partial d}{\partial t}
 = \frac{J_{e}\,M}{n\,F\,\rho}
 - k_{\mathrm{diss}} d ,
\]
where $J_e$ is electron flux, $M$ and $\rho$ are the molar mass and density of the SEI, and $k_{\mathrm{diss}}$ captures dissolution or reconstruction.  
This formalism unifies the diffusion‐limited growth observed in batteries with the dynamic breathing and partial reconstruction characteristic of supercapacitor interphases.\cite{ramadass2002capacity}  
Parameterization from MD or experimental data—ionic conductivity, reaction rate constants, or porosity—enables direct prediction of macroscopic ESR and capacitance decay.

\subsection{Cross-Scale Coupling Strategies}

Multiscale coupling integrates information across five hierarchical tiers:  
(1)~\textit{ab initio} MD or DFT to resolve bond breaking and electronic structure;  
(2)~reactive or ML‐enhanced MD to describe nanometer‐scale nucleation;  
(3)~phase‐field or coarse‐grained dynamics for mesoscale morphology;  
(4)~continuum transport for effective conductivity and mechanical response; and  
(5)~device‐level porous‐electrode or equivalent‐circuit models for performance metrics.  
Cross‐scale demonstrations have revealed universal scaling laws linking SEI thickness, ionic resistance, and energy efficiency in both Li‐ion cells and ionic‐liquid supercapacitors.  
Remaining challenges include rigorous parameter transfer, uncertainty propagation, and maintaining consistent boundary conditions for potential, solvation, and geometry across scales.\cite{bedrov2020multiscale,gabriel2021uncertainty,schomburg2024lithium}

\subsection{Model Benchmarking and Future Directions}

Reliable SEI modeling requires consistent methodologies and quantitative validation.  
While battery modeling benefits from established community benchmarks, comparable datasets for capacitors are scarce.\cite{zhang2023recent,weber2019long}  
Developing shared protocols—standard boundary conditions, force‐field benchmarks, and open reaction databases—would enable direct cross‐comparison and accelerate transferability of results.  
Operando XPS, Raman, and cryo‐TEM provide indispensable validation for both technologies, while ML surrogates trained on combined battery–capacitor datasets may expose universal descriptors such as local electric‐field intensity, surface work function, and solvation number.\cite{he2023capacitive}  
Outstanding challenges include bridging the temporal gap between nanosecond simulations and hour‐scale degradation, expanding reaction libraries beyond carbonates to nitriles and ionic liquids, and quantifying mechanical coupling between SEI formation and electrode deformation.\cite{weddle2023continuum,decaluwe2019open}  
Progress in these directions—anchored by interoperable datasets and open-source workflows—will transform SEI modeling from qualitative visualization to quantitative prediction, ultimately enabling rational interphase design that combines the energy density of batteries with the power capability of supercapacitors.

The multiscale simulation frameworks established in this section provide the theoretical foundation for the machine-learning approaches discussed in Section~6, where data-driven methods accelerate prediction, enable inverse design, and bridge the gap between atomistic accuracy and device-scale applicability.

\section{Machine Learning and Data-Driven SEI Design}

Machine learning (ML) has become a transformative tool for understanding and designing solid–electrolyte interphases (SEIs) across all electrochemical energy-storage systems.  
Whether in lithium‐ion batteries or high‐power supercapacitors, the SEI emerges from coupled electrochemical and chemical processes that span multiple scales—from bond breaking and radical formation to film growth, ion transport, and mechanical reconstruction.  
Because these processes are complex and interdependent, traditional modeling alone cannot capture their full hierarchy.  
Data‐driven approaches, trained on both simulation and experimental results, now provide an adaptive framework that connects quantum accuracy with device‐scale predictability.  
This section highlights how machine‐learning interatomic potentials, spectroscopic analytics, and hybrid multiscale workflows jointly accelerate SEI discovery and design in both batteries and supercapacitors.

\subsection{Machine Learning-Enhanced Potentials for Atomistic Simulation}

Machine‐learning interatomic potentials (MLIPs) bridge the gap between first‐principles fidelity and large‐scale molecular dynamics (MD) simulation applicable to complex interfacial chemistry. \cite{hou2024improving,chen2025redefining}
Traditional classical force fields, while efficient, cannot capture bond formation, charge polarization, or reactive dynamics essential for SEI evolution.  
MLIPs trained on density-functional theory (DFT) data incorporate these phenomena at near-ab initio accuracy but at orders-of-magnitude lower computational cost~\cite{hou2024improving,bin2024perspective,chen2025redefining}.
This paradigm has already proven valuable in both LiPF$_6$/EC battery electrolytes and TEABF$_4$/ACN capacitor electrolytes, indicating that the same ML frameworks can learn transferable interfacial chemistry across different devices\cite{rocken2025enhancing}.

A typical MLIP workflow begins with the generation of a representative dataset encompassing bulk electrolyte configurations, electrode surfaces (including doped carbons, Li metal, MXenes, and defects), solvation shells, and early decomposition products\cite{fletcher2025optimal,wu2025combined}.
Neural‐network or graph‐neural‐network architectures—such as Deep Potential (DeepMD‐kit)~\cite{wang2018deepmd}
, NequIP~\cite{gramatte2024we}, Allegro \cite{maxson2025ms25}, and SpookyNet—predict energies and forces from atomic configurations while enforcing rotational and translational equivariance (E(3) symmetry). \cite{bihani2024egraffbench} 
Such architectures capture the diverse bonding motifs present in SEIs: inorganic fluorides, organic carbonates, nitrile polymers, and metal–oxygen–fluorine species.  
Their data efficiency enables accurate modeling even when trained on relatively small, chemically diverse DFT datasets\cite{chen2020electrolyte,siddiqui2024machine}.

When applied to SEI formation, MLIPs reproduce both the reductive decomposition of solvents and the nucleation of inorganic fragments.  
For example, the same potential can learn EC $\rightarrow$ Li$_2$CO$_3$ reduction pathways relevant to batteries and ACN $\rightarrow$ poly(CN) polymerization seen in capacitors.\cite{wu2025combined,liu2024exploring} 
Because MLIPs combine reactivity with scalability, they capture film nucleation, cross‐linking, and mechanical densification—phenomena inaccessible to fixed‐charge MD.  
They also allow microsecond‐scale trajectories and nanometer‐scale system sizes, bridging the gap between ab~initio MD and continuum models. \cite{liu2024exploring}

Nevertheless, successful deployment demands broad and physically balanced training sets. 
Datasets must cover the full configuration space—bulk, interface, radical intermediates, and decomposed fragments—to avoid extrapolation artifacts.  
In both batteries and capacitors, charge transfer and polarization are key; MLIPs must include charge‐equilibration schemes or explicit electrostatics. \cite{chen2007qtpie}  
Transferability across potential windows and electrode chemistries remains a challenge: models trained solely on bulk or low‐bias conditions can fail under extreme potentials unless continually retrained.  
Active learning, uncertainty quantification, and on‐the‐fly data generation mitigate these issues by automatically sampling configurations where prediction confidence is low. \cite{zaverkin2024uncertainty,wang2024anodes} 

Thus, MLIPs offer a unified platform: they translate electronic‐structure fidelity to mesoscale simulations, enabling predictive studies of SEI formation, growth, and failure across all electrochemical systems.

\subsection{Data-Driven Interpretation of Experimental Characterization}

As the SEI increasingly defines the performance envelope of both batteries and supercapacitors, the ability to predict its chemistry from experimental observations becomes critical. \cite{adenusi2023lithium}  
ML‐based analysis of spectroscopic and imaging data now provides this capability, transforming large, heterogeneous datasets into quantitative chemical maps. \cite{sun2022machine}

High‐dimensional data acquired from X‐ray photoelectron spectroscopy (XPS), time‐of‐flight secondary‐ion mass spectrometry (ToF‐SIMS), Raman and infrared spectroscopy, or hyperspectral electron microscopy contain the full chemical fingerprint of interphases but are notoriously difficult to deconvolute. \cite{tyler2003interpretation} 
Traditional chemometric techniques—principal‐component analysis, multivariate curve resolution—offer limited linear separation and cannot capture nonlinear correlations between spectral features and local chemistry. \cite{chatterjee2018perspective}
In contrast, modern ML methods such as convolutional neural networks (CNNs), autoencoders, and graph‐based architectures excel at uncovering hidden relationships among peaks, textures, and morphologies~\cite{bertz2021deep}.  

In SEI analysis, supervised networks are trained on curated datasets linking spectra or images to ground‐truth composition obtained via XPS deconvolution or ex~situ chemical analysis.  
For example, CNNs applied to ToF‐SIMS maps can distinguish inorganic‐rich from polymer‐rich domains; recurrent networks analyzing operando Raman time‐series can detect transition points from stable to degrading interphases.  
Cross‐domain learning between battery and capacitor datasets—where both share core chemical species (C–O, C–F, LiF, BF$_x$O$_y$)—improves generalization and reduces data requirements.

The benefits are multifold.  
First, ML accelerates screening of electrode/electrolyte combinations by predicting probable SEI compositions from rapid spectral snapshots. \cite{liu2024exploring} 
Second, data‐derived composition vectors can serve as direct inputs for continuum or transport models, eliminating assumptions about generic SEI chemistry. \cite{xu2024challenges}
Third, inverse modeling enables performance‐to‐chemistry inference: given capacitance fade or impedance growth, ML can estimate likely compositional shifts and identify underlying degradation pathways. \cite{zhu2023combined}

To ensure reliability, datasets must include diverse electrode materials (graphite, doped carbons, MXenes, Li metal), electrolyte families (carbonates, nitriles, ionic liquids), and cycling histories.\cite{zheng2024additive,soomro20242d,wang2021situ}  
Label accuracy and balance are critical; semi‐supervised learning, transfer learning, and data augmentation help address data scarcity and imbalance.  
Interpretability tools—such as attention maps and SHAP analyses—link spectral features to specific bonds or chemical groups, ensuring that models retain physical meaning.\cite{xiao2024semi}
Uncertainty quantification should accompany all predictions, particularly when models guide design decisions or automated optimization.

As large, labelled spectral and imaging repositories become available, ML‐based SEI composition prediction will evolve from a diagnostic tool to a predictive engine guiding electrolyte and electrode design in both battery and capacitor technologies. \cite{ng2023machine}

\begin{figure}[htbp]
  \centering
  \includegraphics[width=1\textwidth,page=1]{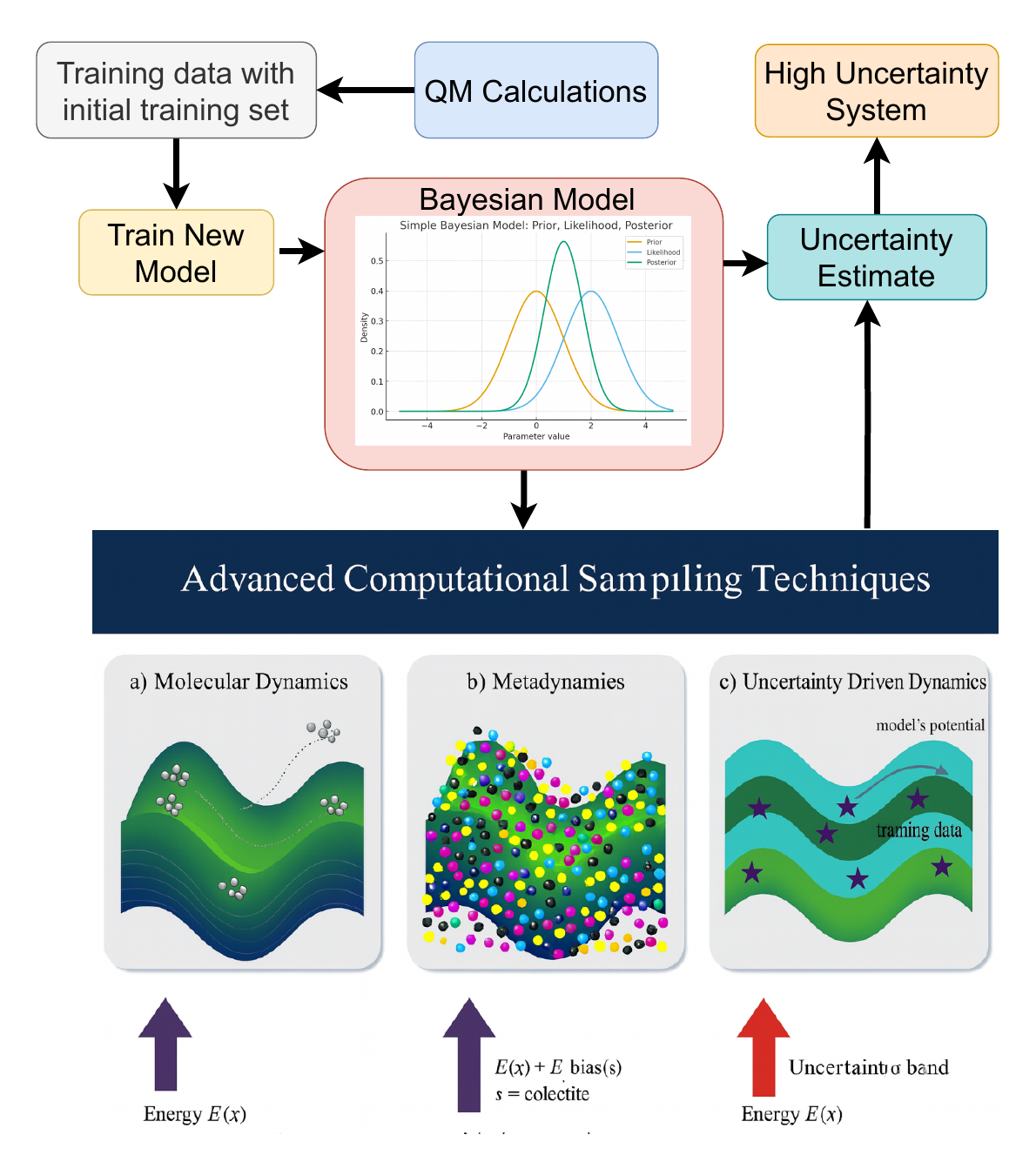}
\caption{
Overview of an iterative workflow for developing interatomic potentials. 
A Bayesian model trained on QM data provides uncertainty estimates that guide additional sampling. 
Advanced techniques—including molecular dynamics, metadynamics, and uncertainty-driven dynamics—are used to explore the energy landscape and refine the potential.
}

  \label{fig:sec_7}
\end{figure}

\subsection{Computational Design of Optimized Interphases}

Machine learning now acts as the connective tissue linking atomistic simulations, materials discovery, and device‐level performance modeling.  
In both Li-ion batteries and supercapacitors, the SEI evolves through reaction–diffusion–mechanical processes spanning orders of magnitude in time and space.  
ML frameworks integrate information from these scales to predict, optimize, and ultimately design interphases with tailored functionality.

\textbf{Accelerating Prediction.}  
Surrogate ML models trained on high-fidelity DFT or MD datasets can instantly estimate SEI descriptors—film thickness, ionic conductivity, dielectric constant, or diffusivity—as functions of electrolyte formulation, potential, and temperature~\cite{wu2025combined}.  
By learning complex nonlinear dependencies, these models bypass direct integration of differential equations, reducing computation while retaining interpretability.  \cite{you2024investigation}
In batteries, such surrogates predict LiF/Li$_2$CO$_3$ layer thickness under cycling; in capacitors, they forecast leakage‐current evolution or impedance rise due to polymeric residue accumulation. \cite{li2024enhancing}
When coupled with continuum aging models, ML surrogates convert atomistic insights into lifetime and stability maps, providing a direct link between nanoscale chemistry and macroscopic performance.\cite{zhang2024liquid}

\textbf{Data-Guided Inverse Design.}  
Beyond prediction, ML enables targeted electrolyte and additive discovery.  
Supervised and Bayesian‐optimization frameworks learn mappings between molecular descriptors (reduction potential, donor number, viscosity) and SEI metrics (composition, resistance, ionic transport). \cite{wang2023recent,li2024ai} 
Once trained, these models can propose new solvent–salt–additive combinations that maximize interphase stability while minimizing electronic leakage.  
Battery‐oriented efforts have already demonstrated automated electrolyte screening; extending these pipelines to capacitors requires inclusion of parameters relevant to EDL structure, dielectric constant, and surface wettability. \cite{ma2025active} 
Feature‐importance analyses reveal mechanistic trends—such as the dominant role of anion reduction potential or solvent polarity—that hold across both technologies. \cite{wang2023recent}

\textbf{Hybrid Multiscale Workflows.}  
The ultimate goal is a closed AI + physics loop where quantum calculations, atomistic MD, continuum models, and experiments continuously inform one another.  
A representative pipeline begins with DFT screening of reduction barriers for solvent and anion fragments; ML‐enhanced MD simulates interphase nucleation and structural evolution; surrogate models extrapolate long‐time growth; and continuum or equivalent‐circuit models predict capacitance fade, ESR increase, or Coulombic inefficiency.  
Experimental feedback—spectroscopic composition or impedance evolution—updates ML models through active learning, ensuring fidelity and uncertainty calibration.  
Such hybrid workflows automate interphase discovery and enable rapid prototyping of stable chemistries for both high‐energy and high‐power devices.

Together, these approaches constitute an ML‐driven multiscale paradigm where data, physics, and experiment converge.  
By linking microscopic descriptors of SEI chemistry (bond energies, solvation motifs) to macroscopic metrics (resistance, capacitance retention), ML transforms SEI research from descriptive observation to predictive and generative design, unifying battery and capacitor interphase engineering.

\subsection{Infrastructure for Reproducible SEI Research}

The foundation of machine‐learning progress is high‐quality, interoperable data.  
While the battery community has begun establishing open datasets for electrolyte decomposition and SEI characterization, comparable resources for supercapacitors remain scarce.  
A unified benchmark repository—spanning both technologies—would dramatically accelerate cross‐fertilization of modeling approaches. \cite{yan2024non,sial2024advancement}

An ideal dataset would record electrolyte composition (solvent, salt, additive, concentration), electrode characteristics (surface area, functionalization, defect density), cycling conditions (voltage range, current density, cycle number), measured SEI properties (thickness, composition from XPS/ToF‐SIMS, ionic/electronic resistance), and device‐level performance (capacitance retention, self‐discharge, ESR).  
Consistent metadata—temperature, humidity, geometry—must accompany every entry.  
Such structured datasets would allow ML models trained on one class of devices to transfer learning to another, enabling joint benchmarking and discovery of universal SEI descriptors. \cite{tan2023structural}

Recent community initiatives~\cite{decaluwe2019open,rajagopal2023data} advocate standardized protocols and open-source repositories for interphase modeling.  
Adapting these practices to the supercapacitor domain—and linking them with battery databases—will create the first truly interoperable SEI dataset ecosystem.  
This infrastructure will enable reproducibility, quantitative benchmarking, and generative design workflows where algorithms autonomously propose optimal electrolyte–electrode combinations for specific stability, conductivity, or mechanical targets.

\subsection{Emerging Paradigms in Interphase Discovery}

Machine learning and data-driven modeling now stand at the frontier of SEI research~\cite{chen2024design}. 
Their integration with physical theory (Section~5) and experimental validation (Section~4) promises a future in which interphase chemistry is not merely characterized but computationally designed.  
By unifying datasets, algorithms, and modeling protocols across batteries and supercapacitors, the community can establish a shared predictive foundation—accelerating discovery of robust, adaptive SEIs that deliver both the high energy of batteries and the high power of capacitors~\cite{nam2024standardized,zhang2026physics}.

The modeling and machine-learning frameworks established in Sections~5 and~6 provide the theoretical and computational tools for SEI prediction and optimization.
Section~7 now translates these insights into practical materials and interface engineering strategies, showing how deliberate control of electrode surfaces, electrolyte chemistry, artificial coatings, and mechanical stabilization enables targeted SEI design in real devices.

\section{Connecting SEI Microstructure to Device Performance}

The architecture and performance of supercapacitors are governed not only by the individual components—electrodes, electrolyte, separator—but by the interplay of their interfaces, most notably the solid–electrolyte interphase (SEI). 
This section explores how the microscopic features of the SEI—its composition, thickness, ionic resistance, uniformity and morphology—translate into macroscopic device parameters such as capacitance, leakage current, equivalent series resistance (ESR), self-discharge, and cycle life~\cite{lokhande2020materials,owejan2012solid,kulova2022electrode}. 
We then examine how these interfacial attributes may be incorporated into device-level models (equivalent circuits, porous-electrode frameworks) to enable predictive links between materials design and performance~\cite{peyrow2023physicochemical,single2019theory}.

\subsection{Microstructural Determinants of Electrochemical Metrics}

The solid–electrolyte interphase (SEI) exerts a decisive influence on the electrochemical performance of supercapacitors by bridging nanoscale interfacial chemistry with macroscopic device behavior. Because the SEI occupies the immediate region between the electrode surface and the electrolyte, its thickness, composition, and uniformity directly regulate ion accessibility, charge‐transfer pathways, and parasitic reactions.\cite{housel2018surface,lokhande2020materials} A thin, ion‐permeable SEI enables rapid ion adsorption and desorption, preserving high specific capacitance and low internal resistance. In contrast, a thick or defective SEI impedes ion migration, reduces accessible surface area, and induces localized ionic starvation—manifesting as lower capacitance and higher equivalent‐series resistance (ESR). Experimental aging studies consistently link capacitance fading and ESR growth to electrolyte decomposition, pore blockage, and interfacial degradation \cite{lokhande2020materials,owejan2012solid,kulova2022electrode}. 

Beyond capacitance, the SEI critically modulates leakage current and self‐discharge. Leakage current in a charged supercapacitor partly arises from unintended Faradaic reactions at the electrode–electrolyte boundary. A stable SEI that is electronically insulating yet ionically conductive suppresses these parasitic currents and minimizes self‐discharge, whereas a heterogeneous or discontinuous interphase promotes ongoing electrolyte breakdown and charge loss \cite{an2016state,liu2023revealing}. From an electrical standpoint, both ionic and electronic resistances contribute to ESR: ion transport through the SEI film and continuity of the conductive network across the electrode–SEI–electrolyte junction determine overall device impedance. As the SEI thickens or becomes more tortuous, ESR rises, diminishing rate capability and power output~\cite{peyrow2023physicochemical,talian2024impedance}. 

These mechanistic correlations translate directly into measurable device‐level outcomes. For example, supercapacitor cells employing carbon‐nanotube electrodes coated with an ultrathin ALD alumina interlayer exhibit markedly lower leakage current and superior capacitance retention after 10\,000 cycles compared to uncoated controls. \cite{lege2023reappraisal} The artificial interphase suppresses electrolyte decomposition and parasitic reactions, maintaining a uniform, stable SEI. Likewise, flexible planar supercapacitors with gel‐polymer electrolytes and cross‐linked binder networks achieve 95\% capacitance retention after 50\,000 bending cycles, illustrating how mechanical stabilization of the SEI enhances durability. \cite{ali2025sodium} In pseudocapacitive hybrid electrodes such as MnO$_2$/graphene, incorporation of a fluorinated electrolyte additive promotes a thin fluorinated SEI, yielding higher rate capability (up to 50\,A\,g$^{-1}$) and lower ESR growth relative to additive‐free devices. \cite{lee2024regulating}
Post‐cycle XPS analyses confirm thinner, homogeneous SEIs enriched in M–F and C–F fragments, directly correlating with improved charge transport and stability. \cite{ji2024trend}

Collectively, these observations establish a coherent interphase–performance link: optimized SEI formation—through artificial coatings, electrolyte or additive design, and mechanical stabilization—enhances ion accessibility, suppresses leakage and ESR growth, and extends cycle life. Treating the SEI as a designed component rather than a passive by‐product thus enables supercapacitors that combine high energy and power densities with exceptional long‐term reliability.\cite{ali2025sodium,cao2022epitaxial}

\begin{figure}[htbp]
  \centering
  \includegraphics[width=1\textwidth,page=1]{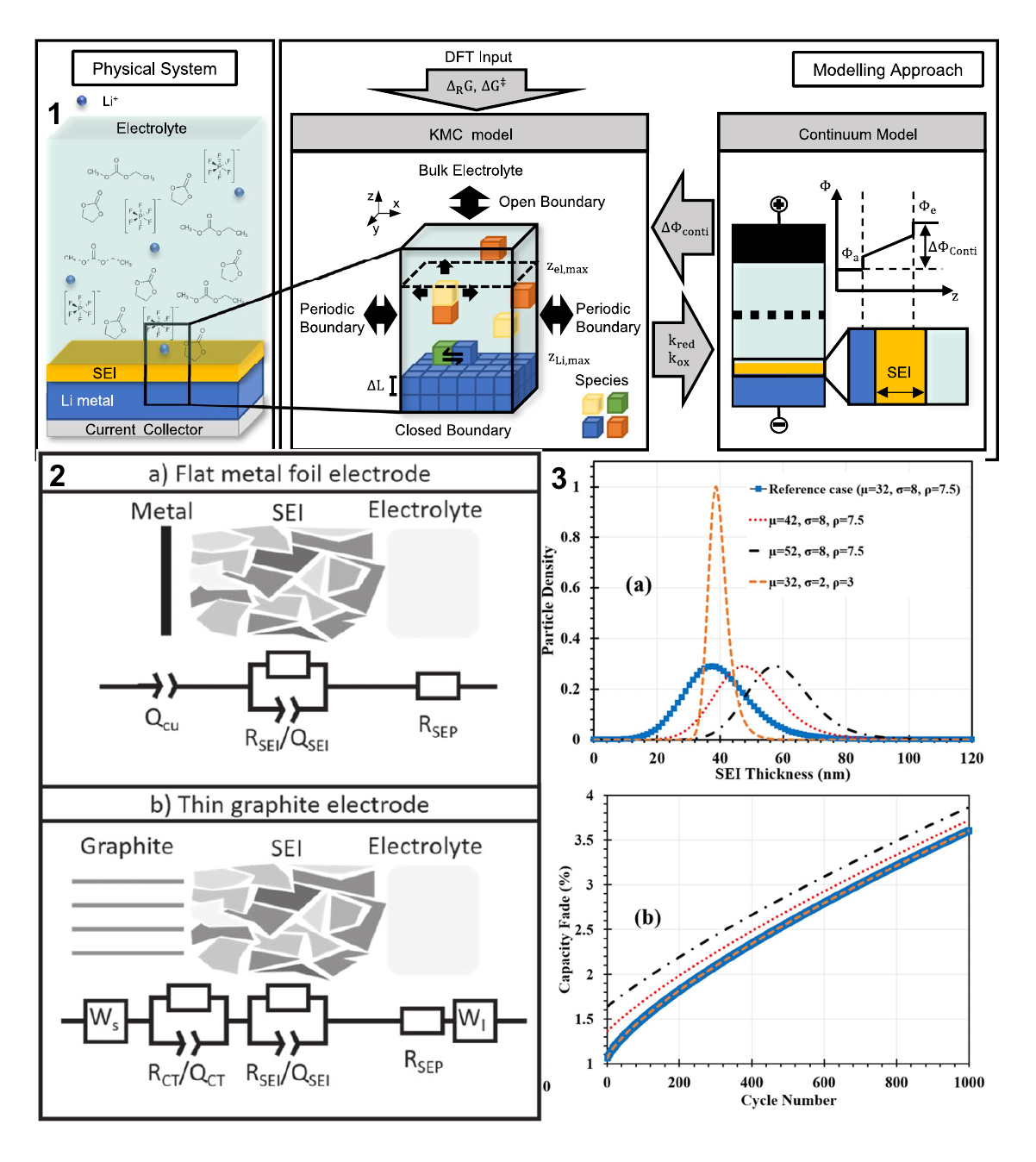}
\caption{1)Schematic overview linking the physical Li–electrolyte system with KMC-based SEI growth modeling and continuum-scale electrochemical transport\cite{wagner2023knowledge}. 
2)Equivalent circuit models for the SEI analysis a metal foil equivalent circuit\cite{morasch2024li}. 3)SEI-thickness-distribution (a) and corresponding capacity fade including the cycle number(b)\cite{tahmasbi2017statistical}
}

  \label{fig:sec_8}
\end{figure}

\subsection{Embedding Interphase Physics in Device Models}

To translate interphase-level understanding into device-level predictive capability, modeling frameworks must embed SEI properties explicitly. Two modelling paradigms dominate: equivalent circuit modelling and porous-electrode (continuum) modelling.

Equivalent circuit models represent the supercapacitor as an ideal capacitor in series and parallel with resistive and reactive elements: typically capacitance (C), equivalent series resistance (ESR), leakage resistance (R\(_{\rm leak}\)), and sometimes a distributed transmission-line component to represent pore impedance. \cite{peyrow2023physicochemical}Within this simplified framework, the SEI may be represented as an additional resistive and capacitive layer: a series resistance and a parallel leakage path reflecting ion transport and parasitic reactions respectively. As the interphase evolves (thickness increases, transport resistance grows, composition changes), the corresponding circuit parameters change. \cite{peyrow2023physicochemical,locorotondc2019modeling}
For example, ESR rises, leakage resistance lowers, and effective capacitance is reduced. Careful calibration of these equivalent circuit elements to measured electrochemical impedance spectroscopy (EIS) data enables linkage of interphase degradation to device-level performance. Practically, characterization of ESR growth, self-discharge current, and capacitance fade over cycling allows reverse deduction of SEI thickening or pore blocking \cite{lee2024regulating,sarode2024solid}.

Porous-electrode models build on more physically realistic descriptions of ionic and electronic transport within electrode architectures. They couple mass transport in electrolyte, ion adsorption at electrode surfaces (double‐layer or pseudocapacitive behaviour), electron conduction in the solid phase, and interfacial kinetics. \cite{peyrow2023physicochemical} In such frameworks, the SEI can be modelled as an interfacial film with defined thickness, ionic conductivity, porosity, and reaction kinetics. \cite{single2019theory} For example, ionic diffusion through pore networks is slowed by an SEI film that reduces pore diameter or increases tortuosity; thereby, the effective ionic diffusivity and active surface area are reduced. \cite{peyrow2023physicochemical} Recent reviews of transport in charged porous media highlight how film layers and pore constrictions modulate ion accessibility and charge dynamics. By embedding SEI parameters into porous electrode models—such as adding an ionic resistance layer, reducing pore cross-section, or adjusting double-layer capacitance per unit area—one obtains predictive time-dependent performance simulations: charge–discharge curves, rate capability envelopes, EIS Nyquist spectra evolution, and cycle life prediction. These models enable bridging between atomic-scale SEI properties (thickness, composition, ionic transport) and system metrics (capacitance retention, ESR drift, self-discharge). Moreover, multi‐physics modelling (coupling electrochemical, thermal and mechanical submodels) can incorporate temperature effects on SEI conductivity, mechanical cracking of the film, and consequent performance degradation. \cite{bedrov2020multiscale,perez2024unraveling}

\subsection{Synergistic Experimental-Computational Platforms}

In modern supercapacitor research, the interphase between electrode and electrolyte no longer remains a mere passive layer but has evolved into a controllable interface through the convergence of in-situ/operando measurement techniques, molecular dynamics (MD) modelling, and machine learning (ML)-based data interpretation. \cite{wang2020electrode} This closed-loop approach begins with high-temporal-resolution experimental measurements such as electrochemical impedance spectroscopy (EIS), electrochemical quartz-crystal microbalance (EQCM), operando Raman or X-ray reflectivity, and cryo-TEM imaging, all of which provide real-time signatures of the solid–electrolyte interphase (SEI) formation, ion transport changes and morphological evolution under cycling. \cite{diddens2022modeling,xiang2025cryogenic} These experimental datasets feed into ML algorithms that rapidly interpret spectral, imaging and impedance features, infer SEI metrics (e.g., thickness, ionic resistance, porosity, heterogeneity) and compare them to predictive models derived from MD. The ML output then informs MD (or ML-surrogate) simulations under the updated interphase state, applied voltage protocols and electrolyte conditions. \cite{yang2023characterization,galvez2018molecular} Simultaneously, device-level performance metrics—such as capacitance drift, leakage current increase and equivalent series resistance (ESR) growth—are monitored to validate or recalibrate the model, thereby creating a feedback loop: experiment → ML insight → simulation → updated experiment.

For supercapacitors, where high rate cycling, thin films and rapid interphase evolution are typical, this feedback loop enables real-time interphase management. For example, when impedance growth accelerates unexpectedly, the ML model can signal that the SEI ionic resistance has reached a threshold, prompting an adjusted cycling protocol (e.g., reduced voltage hold, rest period) or trigger an additive injection. The MD component provides mechanistic insight—how desolvation barriers, pore blockage or ion-film interaction evolve with film composition—thus strengthening the predictive fidelity of the workflow. \cite{li2023accelerated}
This proactive framework transforms the SEI from a passive, post-mortem phenomenon into an actively managed interface.

The power of this integrated loop extends to accelerated lifetime prediction: early‐cycle fingerprints of SEI growth (e.g., rapid increase in ionic resistance or specific Raman shifts) are correlated by ML with long‐term cycle performance, enabling the simulation of interphase evolution under projected usage conditions. This anticipatory capability is especially valuable in high‐power supercapacitors, where ensuring interphase stability over thousands of fast cycles is a prerequisite. By tying simulation parameters to actual device conditions (ion concentration, pore wetting state, morphological change) rather than idealised models, the feedback approach significantly reduces the gap between laboratory insight and real‐world device behaviour.\cite{sheng2021green}

Nevertheless, successful deployment of such in-situ/ML/MD feedback loops demands rigorous data synchronisation (experiment vs simulation time scales), transparency of ML decision-making (avoiding black-box predictions), and sufficiently large and diverse training datasets linking SEI signatures to performance across cycling regimes and chemistries. In spite of these challenges, the closed-loop paradigm stands as a transformative route toward predictive, real-time SEI control and device‐level interphase optimisation in next-generation supercapacitors.\cite{liu2019li}

\subsection{Interphase Engineering for Advanced Device Configurations}

Hybrid supercapacitors—which combine characteristics of both batteries (high energy) and capacitors (high power)—alongside solid-state supercapacitor architectures impose distinctive demands on the solid–electrolyte interphase (SEI).\cite{jin2022better} In battery–capacitor hybrids (such as Li-ion or K-ion insertion anode paired with a capacitive cathode), the SEI must tolerate high current density, large ion flux, and broad voltage swings while enabling rapid ion exchange typical of capacitors. In such systems, the negative electrode often undergoes faradaic insertion and SEI formation similar to battery anodes. \cite{liang2023low,jia2024recent}
Studies on non-aqueous potassium–ion hybrid supercapacitors reveal that formation protocols influence SEI composition (e.g., KF content) and cycling stability, demonstrating the critical role of the interphase in hybrid performance. \cite{majumdar2020journey,xia2021recent} Meanwhile, in solid-state supercapacitors (polymer or ceramic electrolytes), the interphase becomes a “solid–solid” boundary layer that must maintain mechanical and ionic continuity under compression, bending or thermal cycling. Here, the SEI must be ionically conductive, electronically insulating, and mechanically robust—often serving simultaneously as the electrolyte interface and protective interphase.\cite{zhang2024liquid}

From a modelling and engineering perspective, these hybrid and solid-state systems demand tailored SEI strategies. For hybrids, simulation must incorporate ion insertion kinetics, interphase growth under dynamic flux, and coupling between Faradaic and capacitive processes. \cite{ali2019first} For solid-state systems, interphase modelling must include mechanical deformation, adhesion, stress propagation, and film continuity under mechanical load. \cite{zhang2024liquid} The engineering implications are significant: electrodes must be designed to tolerate mechanical strain and rapid flux, while electrolytes must be formulated to produce interphases that resist delamination, preserve ion pathways and mitigate interface degradation. \cite{jia2024li} Accordingly, SEI design in hybrid supercapacitors demands a holistic view integrating insertion chemistry, interphase transport, mechanical resilience and high‐rate operation, enabling devices that deliver both high energy and high power with durability. \cite{jin2022better}

\subsection{Design for Sustainability and End-of-Life Recovery}

Beyond immediate performance improvements, the engineering of the SEI must address long‐term sustainability and recyclability of supercapacitor technologies. \cite{ng2020non} The interphase layer, frequently composed of decomposition products, polymer fragments or engineered coatings, influences device lifecycle, end-of-life behaviour and material recovery. From a materials‐circularity standpoint, SEI design must consider environmentally benign solvents, salts and additives, low‐toxicity coatings, and film formulations that facilitate safe disposal or regeneration. For instance, avoiding high‐toxicity fluorinated additives or unstable polymer residues may reduce environmental footprint and simplify recycling workflows. \cite{capson2020unraveling}

Recycling supercapacitors often requires predictable interphase chemistry: electrodes coated with stable but removable or convertible layers allow more efficient separation of electrode, current collector and electrolyte during end-of-life processing. Engineered SEIs that deposit uniform inorganic salt layers (rather than uncontrolled polymeric debris) enable more consistent recycling behaviour and material recovery. Moreover, the reduction of self-discharge and leakage current through optimized SEI design extends device lifetime and thereby reduces waste associated with early replacement.

From a manufacturing and lifecycle perspective, SEI engineering also influences formation times (with energy savings), cell safety (reduced leakage), and module durability (fewer replacements)~\cite{huang2023salt}. 
By integrating sustainability considerations into interphase design—selecting benign chemistries, designing films that facilitate separation or recovery, and minimising long-term degradation—researchers align high performance with environmental and economic viability~\cite{chen2024design,zhang2024cryo}.
Ultimately, treating the SEI not only as a performance filter but as a sustainability fulcrum ensures that next-generation supercapacitors deliver high energy, high power and high longevity while embodying circular-economy principles.

\subsection{Summary}

This section has demonstrated how SEI microstructure—composition, thickness, ionic resistance, and morphology—directly determines macroscopic device metrics including capacitance, leakage current, ESR, and cycle life.
By embedding interphase parameters into equivalent-circuit and porous-electrode models, and by establishing in-situ feedback loops that couple MD, ML, and experiment, researchers can now predict and control device performance from first principles.
The insights from Sections~2 through~8—spanning fundamental electrochemistry, microscopic mechanisms, experimental characterization, multiscale modeling, machine learning, materials engineering, and device integration—collectively establish a comprehensive framework for SEI science in supercapacitors.
Section~9 builds on this foundation to identify strategic future directions that will transform SEI engineering from an empirical art to a predictive science.

\section{Future Directions and Opportunities}

As the field of solid–electrolyte interphase (SEI) engineering for supercapacitors matures, several overarching opportunities emerge that promise to elevate interphase design from empirical trial‐and‐error to predictive, data‐driven, and sustainable engineering. 
We highlight five strategic frontiers: unification of interphase theory across batteries and capacitors; artificial‐intelligence (AI)–accelerated multiscale simulation for real‐time interface evolution; open databases for SEI chemistries and degradation pathways; coupling quantum‐electronic transport simulations with interfacial models; and design of sustainable electrolytes and eco‐friendly SEI layers for next‐generation energy storage.

\subsection{Unified Theory Across Energy-Storage Modalities}

Historically, SEI research has primarily focused on rechargeable batteries, especially lithium-ion systems, where the SEI at the anode plays a critical role in capacity retention and safety~\cite{gao2019polymer,tan2021growing}. 
In contrast, SEI phenomena in supercapacitors have received only limited attention—despite the fact that electrode–electrolyte interphases in high-power capacitive devices face analogous challenges (e.g., electrolyte decomposition, ion‐transport impediment, interphase mechanical fatigue). 
A unifying theoretical framework that spans both battery and supercapacitor interphases would facilitate cross-pollination of mechanistic insights—such as electron tunnelling, ion‐solvent reorganisation, interphase growth kinetics and mechanical failure modes~\cite{tan2023structural, quan2021unveiling}.

Such convergence requires systematic comparison of interphase chemistries, growth pathways, transport resistances and mechanical behaviours under the distinct operating regimes of batteries (slow charge, deep insertion, large volumetric changes) and capacitors (fast charge/discharge, high ion flux, minimal diffusion depth). 
By leveraging common descriptors (e.g., electron tunnelling barrier, ion-film diffusivity, film fracture toughness), researchers can develop shared modelling tools, standardised measurement protocols and unified degradation metrics. 

\subsection{Machine-Learning-Enhanced Multiscale Modeling}

A second frontier lies in the deployment of artificial intelligence (AI) to accelerate multiscale simulations that capture SEI formation and evolution in real time. 
Recent advances in physics-informed machine-learning workflows have shown that neural‐network potentials and surrogate models can bridge atomic resolution to device timescales by several orders of magnitude~\cite{miksch2021strategies,antonello2023physics}. 
For SEI in supercapacitors, a coherent pipeline might begin with density‐functional theory (DFT) calculations of solvent/salt reduction barriers, feed into molecular dynamics (MD) or reactive‐force-field simulations of early film nucleation, and then leverage ML surrogates to predict film growth, ionic transport and impedance evolution over thousands of rapid charge/discharge cycles.

Incorporating AI into this loop enables adaptive simulations: based on evolving interphase properties (thickness, porosity, composition) the ML model adjusts boundary conditions or cycling protocols and predicts future behaviour before experimental validation~\cite{batra2021emerging}. 
Such “closed‐loop modelling” is particularly powerful in supercapacitors, where fast kinetics and high‐rate protocols dominate and where limited time exists for long‐term ageing experiments. 

\subsection{Community Data Repositories and Standards}

The third strategic opportunity is the creation of shared, open databases linking electrolyte formulation, electrode surface chemistry, SEI composition and performance degradation metrics~\cite{rajagopal2023data}. 
Currently, data in interphase research are fragmented—individual studies report specific electrode/electrolyte pairs and SEI analyses, but lack standard formats, metadata or cross‐system comparability. 
Establishing a comprehensive database would enable machine-learning models to learn from a broader chemical space, quantify trends, identify outliers and propose new interphase chemistries. 

A meaningful SEI database could include parameters such as solvent identity, salt species and concentration, additive content, electrode material and surface functionalisation, SEI thickness/composition (XPS, ToF-SIMS), ionic resistance and capacitance fade across cycling~\cite{rajagopal2023data}. 
This repository would support benchmarking of simulation models, facilitate transfer learning across systems, and catalyse inverse design of interphases. 

\subsection{First-Principles Treatment of Interfacial Charge Transfer}

While much of SEI modelling emphasises ion transport, film growth kinetics and mass-transport limitations, a crucial—but less explored—axis is electron transport through the interphase~\cite{takenaka2021frontiers}. 
Recent quantum‐transport studies reveal that heterogeneous interfaces (e.g., LiF/Li$_2$CO$_3$ combinations) severely influence electron tunnelling barriers and defect‐mediated leakage paths~\cite{zhouunveiling}. 
For supercapacitors operating at high voltages and fast charge/discharge rates, even minor electron conduction through the interphase may trigger continuous electrolyte breakdown, self-discharge or power loss. 

Thus, coupling quantum-mechanical simulations (NEGF, DFT) of interphase materials with MD and continuum models offers a holistic picture where electron tunnelling, ion migration, solvent breakdown and film evolution all interplay~\cite{zhouunveiling}. 
Integrating these quantum calculations into multiscale modelling enables designers to propose SEIs that are not only ionically conductive but also electronically insulating—minimising parasitic reactions and enhancing cycle life. 

\subsection{Green Chemistry and Eco-Friendly Electrolytes}

Finally, the longevity, recyclability and environmental footprint of energy storage systems depend heavily on the interphase. 
While performance has often dominated research priorities, future supercapacitors must align with circular-economy principles. 
This means deploying electrolytes that are non‐toxic, flame-retardant and low-cost, and engineering SEIs that are stable but amenable to safe end-of-life processing~\cite{ju2024self, lin2025decoupling}. 

Sustainable SEI design involves selecting solvents and salts whose decomposition products are benign, coatings and additives that avoid critical raw‐material scarcity, and interphases that enable recycling of electrode materials without extensive chemical breakdown. 
For instance, replacing fluorine‐rich additives with alternative film-formers or employing bio-derived polymer interphases could reduce environmental burden while maintaining performance~\cite{ju2024self}. 
Formation protocols that minimise energy consumption, reduce gas evolution and limit parasitic side‐products will further improve life‐cycle impact. 

By embracing these five strategic directions—unified theory, AI-driven simulation, open data, quantum-informed modelling and sustainability—SEI engineering for supercapacitors can transition from incremental improvements to a generative, predictive design paradigm. 
This transformation will underpin the development of high-energy, high-power, durable and environmentally compatible supercapacitor technologies.

\section*{Concluding Remarks and Outlook}

The scientific understanding of solid–electrolyte interphases (SEIs) has advanced substantially over the past decade, driven by parallel progress in interface-sensitive experimental techniques and increasingly sophisticated computational methods. While traditional electrochemical measurements and ex situ surface analyses laid the foundation of SEI research, they are now complemented by operando spectroscopy, cryogenic microscopy, and multiscale simulation frameworks capable of resolving interfacial dynamics with atomic precision. Together, these developments have transformed the SEI from a phenomenological concept into a quantifiable, chemically heterogeneous interphase governing performance in both batteries and supercapacitors.

In this review, we have focused on the microscopic to mesoscopic mechanisms underlying SEI formation, evolution, and transport, emphasizing the roles of electronic-structure theory, reactive and machine-learning molecular dynamics, continuum models, and device-scale formulations. These approaches, adapted over the years to capture electron transfer, solvation breakdown, and polarization at complex electrode–electrolyte interfaces, now provide access to structural, chemical, and transport properties that were previously inaccessible through experiments alone.

Comparison with experiments provides a critical benchmark for validating multiscale SEI models. In most cases, simulations reproduce the qualitative trends observed in operando XPS, impedance spectroscopy, and cryo-TEM, lending confidence to the underlying assumptions regarding reduction pathways, ion transport, and interphase morphology. Quantitative agreement is increasingly attainable, particularly when simulation parameters are calibrated against ab initio energetics and experimentally measured reaction barriers, though such accuracy often comes at the cost of extensive sampling and high computational demand. The primary benefit of these computational approaches lies in their ability to resolve nanoscale structure, electron redistribution, and decomposition kinetics that remain invisible to most experimental probes. In batteries and supercapacitors alike, these insights have clarified the origins of inorganic–organic layering, elucidated the polarization-controlled onset of electrolyte reduction, and revealed how local electric fields guide SEI nucleation. Moreover, simulations have begun to suggest rational strategies for interface engineering — such as tuning solvent composition, optimizing surface terminations, and exploiting heterogeneous polarization — to stabilize thin, ionically conductive interphases. Although many of these design principles remain challenging to implement experimentally, they provide mechanistic routes for future improvements in interphase stability and device lifetime.

Despite these successes, significant challenges remain for achieving a fully predictive description of SEI formation and evolution across electrochemical technologies. In both battery and supercapacitor systems, the accessible time and length scales of atomistic simulations remain far smaller than those associated with long-term interphase growth, mechanical reconstruction, or electrolyte aging. Bridging these gaps will require tighter coupling between reactive MD, ML-augmented potentials, and continuum models capable of capturing ion transport, stress generation, and dissolution–redeposition dynamics. Methodologically, two major obstacles must be addressed. First, current force fields and ML potentials, while increasingly accurate, still struggle to represent the full chemical diversity of real electrolytes — particularly when multiple solvents, salts, or radical intermediates participate simultaneously in reduction processes. Extending these models to incorporate more explicit electronic descriptors and rare-event pathways will be essential for capturing system-specific reactivity. Second, transferring parameters and boundary conditions across scales remains nontrivial: polarization, solvation, and porosity must be represented consistently from the quantum level to the porous-electrode scale. These issues, combined with the computational cost of reactive simulations and the memory demands of high-resolution mesoscale models, continue to limit the direct simulation of realistic electrode architectures. Overcoming these bottlenecks is therefore central to advancing SEI modeling in the coming years.

In the case of MD-based approaches, substantial progress has been made in representing electrode surfaces and interfacial chemistry, yet important limitations persist. More explicit treatment of electronic structure — particularly localized charge transfer, image-charge effects, and tunneling — would markedly improve predictions of reduction onset and interphase composition, while incorporating bond flexibility and mechanical compliance in electrode models would enable the study of stress-driven SEI cracking and reconstruction during cycling. On the electrolyte side, models have evolved from fixed-charge representations to polarizable and reactive descriptions, yet coupling self-consistent solvent polarization with dynamically fluctuating electrode charges remains technically challenging and is not widely available in mainstream MD engines. Finally, although supercapacitor electrolytes undergo fewer Faradaic reactions than battery systems, rare decomposition events can still influence long-term stability, but modeling these pathways lies beyond the reach of most classical and semiclassical methods. Ongoing efforts in battery SEI modeling — particularly multistate reaction networks, ML-accelerated rare-event sampling, and electronically informed reactive potentials — provide a valuable foundation, and adapting these advances to capacitor chemistries will be essential for capturing degradation and interphase evolution in high-voltage ionic-liquid and organic systems.

However, the most difficult challenges arise when extending SEI modeling to the broader family of emerging electrode and electrolyte chemistries used in both batteries and supercapacitors. While this review concentrated on carbonaceous systems and conventional carbonate or acetonitrile-based electrolytes, next-generation interfaces increasingly involve multicomponent materials whose reactive behavior and electronic structure differ substantially from the paradigms commonly studied. High-voltage spinel oxides, doped carbons, MXenes, and redox-active 2D materials introduce heteroatom-dependent surface states, variable metallicity, and nonuniform polarization, all of which complicate the description of electron transfer and electrolyte decomposition. Similarly, advanced electrolytes — ionic liquids, water-in-salt solutions, fluorinated additives, and hybrid organic–inorganic systems — exhibit complex solvation and reduction pathways that current empirical and reactive potentials do not always capture reliably. Properly representing these systems will require refined models capable of differentiating site-specific charge response, accommodating element-dependent hardness, and resolving situations in which only a subset of atomic species participate in metallic screening. Incorporating these chemical descriptors into constant-potential, reactive, and machine-learning based frameworks remains a demanding task, but doing so is essential for accurately predicting SEI composition and stability across the increasingly diverse set of materials used in modern electrochemical energy-storage devices.

Even greater complexity emerges when considering SEI formation on transition-metal oxides, alloying hosts, and other redox-active surfaces where proton transfer, surface reconstruction, and multielectron redox processes strongly influence interphase chemistry. In aqueous or hybrid electrolytes, these mechanisms couple tightly to solvent reorganization and acid–base equilibria, placing them outside the reliable domain of classical MD and most reactive force fields. Capturing such effects requires more advanced methods — most notably DFT-based ab initio molecular dynamics or finite-field electronic-structure calculations capable of resolving charge transfer, interfacial polarization, and proton-coupled electron transfer at each time step. Initial demonstrations of these approaches have successfully described early reduction events and potential-dependent solvation structure, but their applicability remains limited by system size: ab initio MD can typically simulate only a few hundred atoms, far smaller than the extended interfaces and nanostructured electrodes relevant for practical SEI studies. Machine-learned interatomic potentials trained on large DFT datasets offer a promising route to bridge this gap, providing near-DFT accuracy with classical scalability. However, current models remain at an early stage for complex electrolytes and multicomponent electrode surfaces, and substantial development is still required to ensure transferability across diverse chemistries and operating conditions.

Recent experimental advances have likewise expanded the landscape of SEI-relevant electrolytes. Highly concentrated formulations such as water-in-salt and solvent-in-salt systems introduce new interfacial chemistries, exhibiting nanostructured solvation environments and nonclassical reduction pathways that remain poorly understood at the molecular scale. Their complex organization near charged interfaces, together with experimentally observed coupling to pore topology and surface functionality, presents an open challenge for simulation and demands models capable of capturing collective structuring, fluctuating ion–solvent clusters, and voltage-dependent solvation breakdown. Emerging chemistries — ionic liquids, fluorinated additives, redox-active ionic liquids, and hybrid organic–inorganic electrolytes — offer routes to engineer more robust, self-limiting SEIs, yet their redox behavior requires methods that can treat reactive events and multiple electronic states with comparable accuracy. Conventional MD lacks this capability, motivating the development of electronically informed reactive potentials and nonadiabatic approaches such as surface-hopping to resolve coupled electron–ion dynamics. Taken together, these directions indicate that SEI research across batteries and supercapacitors will remain a vibrant field in the coming decade, with substantial opportunities for advancing microscopic simulation methods, expanding chemical realism, and establishing predictive models that unify interphase behavior across diverse electrochemical systems.

\bigskip


\noindent\textbf{Key challenges and opportunities for future SEI research:}
\begin{itemize}
  \item Extending reactive and ML-certified force fields to cover the full chemical complexity of modern electrolytes and electrode surfaces (mixed solvents, ionic liquids, doped carbons, 2D materials, etc.).  
  \item Building robust multiscale coupling pipelines that consistently preserve solvation, polarization, and porosity across electronic, atomistic, mesoscale, and device levels.  
  \item Implementing constant-potential, electronically informed reactive simulations that are computationally tractable for realistic electrode architectures and long-time evolution.  
  \item Developing open, interoperable datasets and benchmark protocols covering both battery and supercapacitor chemistries, to enable transfer learning, model validation, and community-wide comparison.  
\end{itemize}

\noindent In summary, while the road ahead is challenging, the convergence of advanced experiments, physics-based modeling, and machine learning offers strong prospects for a predictive, technology-agnostic SEI science — one capable of steering future interface design for both high-energy and high-power electrochemical devices.

\bibliography{main}

\end{document}